\documentstyle[prb,aps,preprint,epsfig]{revtex}
\tightenlines

\begin{document}
\draft

\title{
Metallo-dielectric diamond and zinc-blende photonic crystals
}

\author{Alexander Moroz
\thanks{Present address: ESTEC/ESA, Electromagnetics Division,
P.O. Box 299, NL-2200 AG Noordwijk, The Netherlands} 
\thanks{www.amolf.nl/research/photonic\_materials\_theory/moroz/moroz.html}
}

\address{Soft Condensed Matter, Debye Institute,
Utrecht University, Postbus 80000, 3508 TA Utrecht, The Netherlands
}

\maketitle

\begin{abstract}
Diamond and zinc-blende photonic crystals are studied both
in the purely dielectric case and in the presence
of small inclusions of a low absorbing metal.
It is shown that small metal inclusions
can have a dramatic effect on the photonic band structure. 
Several complete photonic band gaps (CPBG's) can open in 
the spectrum, between the $2$nd-$3$rd, $5$th-$6$th, 
and $8$th-$9$th bands. Unlike in the purely dielectric case, 
in the presence of small inclusions of a low absorbing metal
the largest CPBG for a moderate dielectric constant 
($\varepsilon\leq 10$) turns out to be the $2$nd-$3$rd CPBG. 
The $2$nd-$3$rd CPBG is the most important CPBG, because 
it is the most stable against disorder.
For a diamond and zinc-blende structure of nonoverlapping 
dielectric and metallo-dielectric spheres, a CPBG begins 
to decrease with an increasing dielectric contrast roughly at 
the point where another CPBG starts to open--a kind 
of gap competition. A CPBG can even shrink to zero when 
the dielectric contrast increases further. Metal inclusions 
have the biggest effect for the dielectric constant
$\varepsilon\in [2,12]$, which is a typical dielectric 
constant at near infrared and in the visible for many materials, 
including semiconductors and polymers. 
It is shown that one can create a sizeable and robust 
$2$nd-$3$rd CPBG at near infrared and visible wavelengths
even for a photonic crystal which is composed of more than 97\% 
low refractive index materials ($n\leq 1.45$, i.e., 
that of silica glass or a polymer). 
In the case of silica spheres with a 
silver core, the $2$nd-$3$rd CPBG opens for a metal-volume fraction 
$f_m\approx 1.1\%$ and has a gap width to midgap frequency ratio 
of $5\%$ for  $f_m\approx 2.5\%$. Within the $2$nd-$3$rd CPBG of 
$5\%$, absorption remains very small ($\leq 2.6\%$ once the CPBG 
is centered at a wavelength $\lambda\geq 750$ nm), which  should 
be tolerable in most practical applications. The metallo-dielectric 
structures display a scalinglike behavior, which makes it possible 
to scale the CPBG from microwaves down to the ultraviolet  wavelengths. 
Aluminum, copper, and gold cores yield almost identical results,
provided that sphere radius $r_s\geq 250$ nm. For $r_s< 250$ nm the 
results for different metals can be increasingly different with 
decreasing $r_s$, nevertheless, qualitative features remain the same.
These findings open the 
door for any semiconductor and polymer material to be used as genuine
building blocks for the creation of photonic crystals with a CPBG 
and significantly increase the possibilities for experimentalists to 
realize a sizeable and robust CPBG in the near infrared and in the visible. 
One possibility is a construction method using optical tweezers, 
which is analyzed here.
\end{abstract}
\date{March 15, 2002}

\narrowtext


\section{Introduction}

Photonic crystals are structures with a periodically modulated 
dielectric constant. In analogy  to the case of an electron moving in 
a periodic potential, light propagating in a photonic crystal 
experiences multiple scattering leading to the formation of Bloch waves 
and photonic band gaps \cite{By,Y,Souk}.
If the band gap persists for both polarizations and all 
directions of propagation one speaks of a complete 
photonic band gap  (CPBG) \cite{HCS,YGL}.  Light with
frequencies within a CPBG is totally reflected since it
cannot propagate inside the crystal. Yet light can propagate
through waveguides carefully designed within such a photonic crystal,
even when such a waveguide has sharp bends \cite{MCK}.
In the last decade, photonic crystals have enjoyed 
a lot of interest in connection with their possibilities to guide light
and to become a platform for the fabrication of photonic
integrated circuits \cite{Souk,MCK}. There is a  common belief that, 
in the near future, 
photonic crystals systems will 
allow us to perform many  functions with light that ordinary 
crystals do with electrons. 
In addition to numerous potential 
technological applications (filters, optical switches, superprisms,
cavities, etc  \cite{Souk}), photonic crystals
also promise to become a laboratory for testing fundamental  processes
involving interactions of radiation with matter under 
novel conditions \cite{By,Y}.
The presence of a CPBG  causes dramatic changes in the local density
of states, which offers the possibility 
to  control and engineer the spontaneous  emission of embedded 
atoms and molecules \cite{By}.

Unfortunately,  technological  difficulties in fabricating  
CPBG structures rapidly increase
with decreasing wavelength for which a CPBG is required.
Despite the research activities of a large number of experimental 
groups, achievement of a CPBG below infrared (IR) wavelengths 
for both two- and three-dimensional
(3D) photonic structures is still elusive, mainly 
because the required dielectric
contrast $\delta$ to open a CPBG is rather high \cite{MS}. 
Here $\delta=\varepsilon_{max}/\varepsilon_{min}$, where
$\varepsilon_{max}$ ($\varepsilon_{min}$) is the dielectric constant
of a material component, used to fabricate a photonic crystal,
with the largest (smallest) value of $\varepsilon$.
For the best geometries
$\delta\approx 5$ is required  \cite{HCS,AMS}. Already this threshold
value of $\delta$ excludes the majority of  semiconductors and other 
materials, such as (conducting) polymers, 
from many  useful photonic crystal applications.
However, for applications one needs a sufficiently large CPBG to leave a margin
for gap-edge distortions due to omnipresent defects \cite{ZLZ}. 
Let us define $g_w$  as the gap width to midgap 
frequency ratio, $\triangle\omega/\omega_c$. Then in order to achieve 
$g_w$ larger than  $5\%$, $\delta\gtrsim 9.8$ and 
$\delta\gtrsim 12$ is required for a diamond \cite{AMS}
 and face-centered-cubic (fcc) structure \cite{MS},
respectively. This leaves only a couple of materials
for photonic crystals applications at near infrared and
optical wavelengths \cite{Lev}.

Surprisingly enough, it will be demonstrated here that there 
is a way to create a sizeable and robust CPBG in the 
near infrared and in the visible even for a photonic crystal
which contains more than 97\% of a material with the refractive 
index below $n=1.45$, i.e., that of silica glass or a polymer.
The trick is to place small
inclusions of a low absorbing  metal in the right dielectric structure. 
It turns out that the effect of small inclusions of a low
absorbing metal on the photonic band-gap structure is very strong
for a parent diamond dielectric structure. 
In the latter case, the actual metal-volume fraction $f_m$ needed to 
open a CPBG of more than $5\%$ depends on the 
available material dielectric constant $\varepsilon$ and
can be kept below $1\%$ for $\varepsilon_s=4$ 
(Fig. \ref{gwAgr80xx} below). Given a metal-volume fraction $f_m$,
the strongest effect on CPBG was found when both spheres in the
lattice primitive cell contained identical metal cores.
Surprisingly, the inclusions have the biggest effect for the
dielectric constant $\varepsilon\in [2,12]$, which is
a typical dielectric constant at near infrared
and in the visible for many semiconductors and polymers.
Given the desired gap width to midgap 
frequency ratio of $5\%$, the smallest absorption 
was found when only one of the two spheres in the
lattice primitive cell contained a metal core, i.e.,
for a zinc-blende structure, even though the
required $f_m$ was typically twice as large as that for 
a diamond structure. This result suggests that, given the metal 
filling fraction $f_m$, absorption is reduced
due to a reduction of near-field electromagnetic 
energy transfer between the metal cores with an increased
separation of the metal islands in the structure.
For a zinc-blende structure, absorption within 
a CPBG of $5\%$ can be kept below  $2.6\%$ once the CPBG 
is centered at a wavelength $\lambda\geq 750$ nm. 
The results on the absorption are by far the best which
have been demonstrated for a 3D metallo-dielectric structure
with a CPBG. They are an order of magnitude better than for 
the case of an fcc lattice of metal coated spheres \cite{WCZ}.
Our results on the absorption compare well to the best results for 
one-dimensional and two-dimensional
metallo-dielectric photonic crystals which show absorptance of 
$\approx 1\%$ and $\approx 3\%$, 
respectively, at $\lambda\approx 600$ nm \cite{AMs}. 
Photonic band-structure calculations also
revealed a surprising scalinglike behavior of our metallo-dielectric
diamond and zinc-blende structures, which is only intrinsic
to purely dielectric structures. The  scalinglike behavior 
means that once a CPBG is found, the CPBG can be open for any wavelength, 
simply by scaling all the sizes of a structure--an extremely
useful property from a practical point of view.

A metal-core dielectric-shell sphere morphology seems to play an
essential role in the effect of small metal inclusions on photonic 
band structure. If the same volume of a metal
is spread homogeneously within the spheres, no CPBG is found.
Similarly, a diamond structure of small metal nanospheres embedded 
in a dielectric matrix shows a much smaller CPBG and much higher
absorptance for a comparable $f_m$ than either a diamond or a zinc-blende
close-packed structure of metal-core dielectric-shell spheres in air.
On purely experimental grounds, only the case of spheres 
with a metal core is  investigated here. Usually, a metal shell 
around a dielectric core is formed by an aggregation of small metallic
nanoparticles. The shell has to be around 20 nm thick before it becomes 
complete \cite{GAvB}.  With an emphasis on photonic structures in the near
infrared and in the visible, the 20 nm shell thickness then would mean 
a rather high threshold value of the metal filling fraction $f_m$
(of the order of $5\%$). On the other hand, it is much easier to tune 
the metal filling fraction $f_m$ from zero to a few percent by 
coating small metal nanoparticles with a dielectric in a 
controlled way \cite{MGM}. Moreover, a dielectric shell is necessary 
to prevent aggregation of the metallic particles by reducing 
the Van der Waals forces between them. In the latter case, a 
coating of roughly $20$ nm is required. 
Having metal cores also means that an active light-emitting 
dielectric material (semiconductor or polymer) is not isolated 
as in the case of metal shells. Hence crystals of metal@dielectric
particles can perform  more easily some useful optoelectronic functions.

Before proceeding with the photonic band-structure
calculation of metallo-dielectric diamond and zinc-blende structure of
spheres, it was necessary to recalculate the photonic band 
structure of the diamond lattice of dielectric spheres. The 
previous plane wave method (PWM) calculations of the photonic band 
structure of diamond \cite{HCS} and zinc-blende \cite{SBMc} lattices 
of dielectric spheres were found to be plagued by large errors. Therefore, 
after describing the method of our calculation  in Sec. \ref{sec:dmdbench},
we discuss and analyze the differences with respect to earlier 
photonic band-structure calculations for a diamond lattice
of dielectric spheres by Ho, Chan, and Soukoulis \cite{HCS}.
The effect of small inclusions of a low absorbing metal on the photonic 
band structure of a diamond and a zinc-blende structure 
is investigated in Sec. \ref{sec:doped}. 
In Sec. \ref{ssec:xx} the dependence of CPBG's on the metal
volume fraction $f_m$ is investigated.
The dependence of gaps on either the shell 
dielectric constant in the diamond case, or on the dielectric constant 
of the dielectric sphere in the zinc-blende case, is summarized in 
Sec. \ref{ssec:shell}.
A surprising scalinglike behavior of the metallo-dielectric
structures is then discussed in Sec. \ref{ssec:sca}. Reflectance, 
transmittance, and absorptance of light incident on the metallo-dielectric
structures is dealt with in  Sec. \ref{ssec:rta}. Most of the
results will be shown for the case of silver cores. However,
as discussed in Sec. \ref{ssec:other}, where aluminum, copper, and gold
cores are examined, the results so obtained
are almost insensitive to the type of metal used, provided that
the sphere radius is larger than $\approx 250$ nm. 
If the sphere radius is smaller, most of the results are still 
qualitatively valid.
We then end up with a discussion 
in Sec. \ref{sec:disc} and conclusions in Sec. \ref{sec:conc}.

\section{Benchmarking diamond structure of dielectric spheres}
\label{sec:dmdbench}
Indisputably, one of the hallmarks in the study of
photonic band gap structures is an article
by Ho, Chan, and Soukoulis \cite{HCS}. It opened a way
for the fabrication of the first photonic structure
with a complete photonic band gap (CPBG) \cite{YGL} and advanced
the field considerably. Its main conclusion
is that, regarding a CPBG, a diamond structure of (overlapping) spheres
fares much better than a simple fcc structure: 
(i) a CPBG opens between the $2$nd and $3$rd bands 
(the $8$th-$9$th bands for an inverted fcc structure), 
and, consequently, is much more stable against disorder \cite{ZLZ}, 
(ii) the threshold value of the dielectric contrast 
$\delta$ to open the CPBG is $4$ ($8.2$ for an 
inverted fcc structure \cite{MS}),
and (iii) the CPBG 
is significantly larger ($15\%$ and $5\%$, for the respective
diamond and fcc close-packed lattices of spheres with a dielectric 
contrast $\delta=12.96$). 

Results of our calculations  are summarized in Figs. \ref{diamcosta},
\ref{rgwdmd}. They were obtained using the photonic 
Korringa-Kohn-Rostocker (KKR) method \cite{MS,WZY,Mo} 
(see also Appendix \ref{app:kkr}). 
The KKR method can be used for scatterers of arbitrary shape 
\cite{WM,BGZ} and is optimized for lattices of spheres. 
The photonic band gap
structure was calculated with the value of the angular-momentum
cutoff $l_{max}=9$. Stability of the Ewald summation was thoroughly
checked. Convergence of the method with increasing $l_{max}$ is 
demonstrated in Fig. \ref{dmdconv}. 
The two lowest bands in Fig. \ref{diamcosta} (but not the
higher ones) agree well with those calculated in Ref. \cite{SBMc}. 
In agreement with Refs. \cite{HCS,WZY}, no CPBG was found
for the case of a diamond lattice of nonoverlapping air spheres 
in a dielectric.
However, for the inverse case of a diamond lattice 
of nonoverlapping dielectric spheres in air,
the earlier results of Ho, Chan, and Soukoulis \cite{HCS} 
on the photonic
band structure were found to be nonconverged. Take, for example, the 
test case of a close-packed diamond lattice of nonoverlapping dielectric
spheres with the dielectric constant $\varepsilon_s=12.96$ in air 
(Fig. \ref{diamcosta}).
A nonconvergence of the results of \cite{HCS} is already 
witnessed by an unusually large deviation  of the effective 
refractive index $n_{e\!f\!f}$, as calculated  from the 
band structure in the L direction,  
from $n_{e\!f\!f}^{M\!G}=\sqrt{\varepsilon_{e\!f\!f}}$, 
where $\varepsilon_{e\!f\!f}$ is calculated by the Garnett 
formula \cite{MG},
\begin{equation}
\varepsilon_{e\!f\!f} \approx \varepsilon_h \,
(1+ 2\, f\alpha)/(1- f\alpha),
\label{maxgsp}
\end{equation}
which usually provides a very good fit to $n_{e\!f\!f}$. 
Here $f$ is the sphere volume fraction ($f=0.34$ for closed-packed
diamond and our zinc-blende lattices of spheres) and 
$\alpha$ is a sphere polarization factor
[$\alpha=(\varepsilon_s-\varepsilon_h)/(\varepsilon_s+2\varepsilon_h)$, 
where $\varepsilon_s$
and $\varepsilon_h$ are the dielectric constant of sphere and
of host medium, respectively].
In our case, the effective refractive index
$n_{e\!f\!f}=1.604$ is only $10\%$ larger than $n_{e\!f\!f}^{M\!G}=1.456$. 
For comparison, the upper edge of the bands at
the L point in Ref. \cite{HCS} seems to be almost $20\%$ lower than in our
case, resulting in more than a $30\%$ deviation from
$n_{e\!f\!f}^{M\!G}$. 
Not surprisingly, 6 years after Ref. \cite{HCS} the PWM (see Fig. 4 
of Ref. \cite{SBMc}) yielded for  the same 
test case a $2$nd-$3$rd CPBG width of only $\approx 8\%$, half of
that presented in Ref. \cite{HCS}. 
In the meantime, S\"{o}z\"{u}er, Haus, and Inguva \cite{SHI}, p. 13 971, 
again using the PWM,
argued that the $2$nd-$3$rd CPBG is less than 
$3.5\%$, and they called for a recalculation
of their results using more efficient methods, 
such as the photonic KKR method.

For a diamond lattice 
of nonoverlapping dielectric spheres, i.e., for
sphere filling fraction $f_s$ varying from $0$ to the 
close-packed case $f_{cp}=0.34$, 
two CPBG's can occur simultaneously (see Fig. \ref{diamcosta}), 
between the $2$nd-$3$rd bands, and, as in an inverted fcc case, 
between the $8$th-$9$th bands \cite{AMS}.
According to Fig. \ref{rgwdmd}, the photonic band structure of a 
diamond lattice of nonoverlapping dielectric spheres shows a much
more complex behavior of CPBG's than an fcc lattice.
For an fcc lattice of air spheres in a dielectric, the calculated
gap width to midgap frequency ratio $g_w$ of the $8$th-$9$th
CPBG is monotonically increasing with 
the increasing dielectric contrast
$\varepsilon_h/\varepsilon_s$ and shows a saturated behavior 
for different filling fractions (see Fig. 4 in Ref. \cite{MS}).
A remarkable feature of a diamond lattice 
of nonoverlapping dielectric spheres is a kind of {\em gap competition}.
According to Fig. \ref{rgwdmd}, a CPBG first increases
with the dielectric contrast $\varepsilon_s/\varepsilon_h$, 
but, at some dielectric contrast shortly before another CPBG starts to open,
the CPBG begins to decrease and it may eventually
disappear. As the dielectric contrast $\varepsilon_s/\varepsilon_h$
increases further, the first CPBG may reappear again, with its reappearance
being signaled by a decrease of the second CPBG.
Fig. \ref{rgwdmd}
illustrates that kind of behavior for $f_s=0.17$ and $f_s=f_{cp}$. 
Contrary to previous calculations \cite{HCS}, the $2$nd-$3$rd CPBG is 
found not to be the dominant one. It
satisfies the bound of $3.5\%$ by S\"{o}z\"{u}er, Haus, and 
Inguva (see Ref. \cite{SHI}, p. 13 971) not only
for $\varepsilon_s\approx 13$ but apparently for all $\varepsilon_s$.
For $f_s=f_{cp}$, the $2$nd-$3$rd CPBG only persists for 
$\varepsilon_s\in [5.2, 16.3]$ and does not exceed $2.3\%$ 
(for $\varepsilon_s=9$). The $2$nd-$3$rd CPBG is then absent 
till $\varepsilon_s=40$, where it reappears again.
We disagree with Ref. \cite{SHI} that the $2$nd-$3$rd
CPBG vanishes for $f<f_{cp}$. 
It is true that, for $\varepsilon_s=12.96$, the $2$nd-$3$rd
CPBG  closes already for $f_s=32\%$, i.e., for $f_s$
only $2\%$ smaller than the close-packed case. However, with increasing
$\varepsilon_s$ the $2$nd-$3$rd CPBG can appear at smaller 
and smaller filling fractions. For instance,  for $\varepsilon_s\approx 20$
the $2$nd-$3$rd CPBG begins to 
open for $f_s=17\%$,  but again its $g_w$ does not exceed $2.9\%$,
satisfying the $3.5\%$ bound \cite{SHI}. 
The dominant CPBG is the $8$th-$9$th CPBG. For 
$\varepsilon_s=12.96$, the $8$th-$9$th CPBG persists down to $f_s=4\%$.
For $f_s=17\%$ and $\varepsilon_s=12.96$ it can reach $12\%$,
and it reaches its maximum of $\approx 14\%$ for $\varepsilon_s=15.5$.
For $f_s=f_{cp}$ it reaches a maximum of 
$\approx 14.7\%$ for $\varepsilon_s=35$.
However, the threshold value of $\varepsilon_s$ for its opening is
$7.9$, comparable to that  for an inverted fcc lattice \cite{MS}.
Hence, regarding a CPBG, a diamond lattice 
of nonoverlapping dielectric spheres fares
still better than an fcc structure of air spheres 
in a dielectric \cite{MS}, but the advantage is much smaller 
than has been previously thought.

An argument to persuade the reader about the correctness of 
our result is that they can be reproduced, up to
some minor differences, by the PWM based 
MIT {\em ab initio} program \cite{JJ}. However, unlike the case of 
a simple lattice
 (one scatterer per lattice primitive cell)
\cite{MS}, for the case of a diamond lattice of dielectric spheres, 
even when using the PWM based MIT {\em ab initio} 
program \cite{JJ}, 
one has to take a much larger number of plane waves than expected 
to reach a convergence comparable with the photonic KKR method.
As shown by Megens \cite{Meg},
to reach convergence of the photonic band structure of a diamond 
lattice of dielectric spheres within
$1\%$, the  number of plane waves  $N_c$ has to exceed  $32\, 768$ 
(cf. Ref.  \cite{HCS}) and still an extrapolation
$N\rightarrow \infty$ \cite{SHI} has to be performed. Hence, the
$N\rightarrow \infty$ extrapolation \cite{SHI} might also be 
necessary to ensure a convergence for the MIT {\em ab initio} 
program \cite{JJ} when modeling
systems with sharp material discontinuities for other 
complex lattices. 
Note that $N_c$ is two orders of magnitude  larger than 
$N_{hcs}=339$ which reproduces
the results of \cite{HCS} for the whole range of filling 
fractions $0\leq f\leq 1$ \cite{Meg}. It turns out that
some other results computed 
in the early days of the plane-wave expansion method can be 
imprecise and plagued by large errors also, an example being  
Ref. \cite{VP}.

\section{Effect of metal inclusions on photonic band structure}
\label{sec:doped}
In this section, the effect of small inclusions of 
a low absorbing metal on the photonic band-gap structure for 
a parent close-packed
diamond and zinc-blende dielectric structure of spheres is investigated.
The close-packed case of spheres in air was chosen because this 
is one of the most probable cases to be fabricated either 
by ``do-it-yourself organization''  \cite{AvB} or by a  microrobotic 
technique \cite{SLM}. At the same time, even when starting
from colloidal suspension \cite{BRW}, one expects  to use dried structures 
in air for practical applications which
will most probably consist of touching spheres (close-packed structures).
A metal core will be placed inside either both dielectric spheres
in the lattice primitive cell, or only one single sphere 
in the lattice primitive cell. In the former case, the resulting
structure will be a diamond structure, whereas in the latter case
the structure becomes a zinc-blende structure. In the following
we only consider the special case of zinc-blende structures of spheres,
when both spheres in the lattice primitive cell have identical radii.

To reach convergence of bands with $1\%$ precision in the 
metallo-dielectric case, the photonic KKR method \cite{MS,Mo,AM2} 
with the angular momentum cutoff parameter $l_{max}=8$ was used. 
The KKR method is best suited to deal with highly dispersive 
scatterers. Computational time with and without the dielectric 
constant dispersion is the same and only depends on the angular 
momentum cutoff parameter (see also Appendix \ref{app:kkr}). 
As in earlier work \cite{WCZ,AM2,ZLW}, only the real
part of the material dielectric constant was used when the
photonic band structure was calculated. Reflectance, transmittance, 
and absorptance were calculated by adapting available computer code 
\cite{YSM}, which is based on the layer photonic KKR (LKKR) 
method \cite{Mod}. In the latter case, both the real and imaginary 
part of the material dielectric constant were used. The dielectric 
constant of a metal was chosen according to Ref. \cite{Hand}. 
Above the vacuum wavelength of $\approx 1900$ nm, Palik's data for 
aluminum, copper, and gold are no longer available. In this case a 
Drudelike fit to a metal dielectric constant was used with 
parameters determined by Ordal {\em et al}.\cite{OLB} (Ordal's parameters 
has also been used by El-Kady {\em et al}.\cite{KSB} to fit a 
metal dielectric constant from far infrared down to optical wavelengths.) 
The same procedure for the band structure and absorption calculations 
has also been employed in an earlier work on metallo-dielectric 
structures \cite{WCZ,AM2,ZLW}. 
It is true that a complex dielectric constant turns Bloch
eigenvalues into complex resonances in the lower-half of
the complex frequency plane \cite{TMC}. However, it can be 
shown that the imaginary part of a low absorbing metal has only 
little influence on the projections of the complex resonances on 
the real  frequency axis\cite{KMP1}. This can also be shown by a direct
comparison of the ``projected" band structure against 
reflection and transmission calculations (see Sec. \ref{ssec:rta}). 
Note in passing that a comparison of theory and experiment using 
the photonic KKR methods has so far been more than 
satisfactory\cite{ZLW,VMB}.
 
Photonic band-structure calculations demonstrated that 
several CPBG's can open in the spectrum. Depending on the 
metal filling fraction $f_m \leq 10\%$ and a typical 
dielectric constant $\varepsilon\in(1,20]$ of a dielectric at 
near infrared and in the visible, 
a diamond structure can have CPBG's between the $2$nd-$3$rd and
$8$th-$9$th bands, whereas a zinc-blende structure
can have CPBG's between the $2$nd-$3$rd, $5$th-$6$th, 
and $8$th-$9$th bands \cite{SBMc}. 
Figure \ref{dmdeps12r80T75} shows the photonic band structure for 
a close-packed zinc-blende lattice of spheres in air.
One of the two spheres in the primitive lattice cell
is a silver core - $n_s=1.45$ (silica) shell sphere with
$r_c/r_s=0.75$ ($f_m=7.2\%$), whereas the other is a homogeneous 
dielectric sphere with $\varepsilon=12$
of the same radius $r_s=80$ nm. The frequency is in units [c/A], as in 
Fig. \ref{diamcosta}. By comparing Fig. \ref{diamcosta} and 
Fig. \ref{dmdeps12r80T75} it is obvious that as the result of the metal 
inclusions much larger $2$nd-$3$rd and $8$th-$9$th CPBG's open.
There is also an additional CPBG  between the $5$th and $6$th bands.
Metal inclusions seem to align band edges at the L and W points
and the degenerate band edge of the $3$rd and $4$th bands at the
L point raises above the band edge of the $3$rd band at the U point.

In the following subsection, the dependence of the gaps on the metal-volume 
fraction is investigated. The dependence of gaps on either the shell 
dielectric constant in the diamond case, or on the dielectric constant 
of the dielectric sphere in the zinc-blende case, is summarized in 
Sec. \ref{ssec:shell}.
A surprising scalinglike behavior of the metallo-dielectric
structures is then discussed in Sec. \ref{ssec:sca}. Reflectance, 
transmittance, and absorptance of light incident on the metallo-dielectric
structures is dealt with in Sec. \ref{ssec:rta}.

\subsection{Dependence of gaps on the metal-volume fraction}
\label{ssec:xx}
Photonic band-structure calculations revealed a strong effect
of metal inclusions on the CPBG between the $2$nd and $3$rd bands.
Given a metal-volume fraction $f_m$, the largest $2$nd-$3$rd CPBG opens 
for a close-packed diamond lattice of metal core-dielectric 
shell spheres in air (Fig. \ref{gwAgr80xx}), followed by a 
close-packed zinc-blende structure (Figs. \ref{gwTxxsioa},  \ref{gw300Txx}). 
Unlike in the purely
dielectric case, which has been discussed in Sec. \ref{sec:dmdbench}, 
in the presence of small inclusions of a low absorbing metal the dominant
CPBG for a diamond lattice is that between the $2$nd and $3$rd bands (Figs. 
\ref{gwdmdrf5eps}, \ref{gwdmdrf5r300}). The $2$nd-$3$rd CPBG is the most 
important CPBG, because it is the most stable against disorder \cite{ZLZ}.
In the zinc-blende case,
the $2$nd-$3$rd CPBG is also much larger than in the purely dielectric case 
and is a dominant CPBG for $\varepsilon\in (1,9]$ (Figs. \ref{gwT75epsa}, 
\ref{gwr80T75eps}). For higher $\varepsilon$ the dominant CPBG becomes
the $5$th-$6$th CPBG (Figs. \ref{gwT75epsa}, \ref{gwr80T75eps}).
Photonic band-structure calculations revealed several remarkable 
features  of the diamond and zinc-blende metallo-dielectric structures.
First, a strong increase of the $2$nd-$3$rd CPBG with $f_m$, 
once a threshold value of $f_m$ is reached. 
Figure \ref{gwAgr80xx} shows that if the silver cores are placed inside
spheres with a refractive index
$n_s=1.45$ (silica) and a radius of $80$ nm, the $2$nd-$3$rd CPBG below 
$600$ nm opens for $f_m\approx 1.1\%$ and reaches
$5\%$ already for $f_m\approx 2.5\%$. If the sphere refractive index 
$n_s$ increases
further and approaches the threshold refractive index contrast of 
$\approx 2.3$, for which the  $2$nd-$3$rd CPBG of the parent 
diamond structure of non-overlapping dielectric spheres 
begins to open (see Fig. \ref{rgwdmd}),
the required metal $f_m$'s rapidly decrease.
For $n_s=2$ (ZnS) spheres with a silver core, the
$2$nd-$3$rd CPBG begins to open already for $r_c/r_s=0.13$ 
($f_m \approx 0.07\%$)  and reaches $5\%$ already for 
$r_c/r_s \approx 0.26$ ($f_m\approx 0.6\%$). Figure \ref{gwTxxsioa} shows
that in the zinc-blende case one needs almost twice as large $f_m$ 
as in the diamond case to obtain a comparable $2$nd-$3$rd CPBG.
For example, to obtain the $2$nd-$3$rd CPBG of $5\%$ 
in the case when silver cores are only placed inside
a single $n_s=1.45$ (silica) sphere in the lattice primitive
cell, one needs $f_m\approx 5\%$. Figure \ref{gw300Txx} shows the effect
of the dielectric sphere morphology on the  $2$nd-$3$rd CPBG,
which is either a $n_c=2$ (ZnS) core - silica shell dielectric sphere with 
fixed $r_c/r_s=0.60$ (dashed line) or a homogeneous
silica sphere (solid line). The other sphere in the
lattice primitive cell is a silver core - $n_s=1.45$ (silica) 
shell sphere.

\subsection{Dependence of gaps on the dielectric constant}
\label{ssec:shell}
Figures \ref{gwdmdrf5eps}, \ref{gwdmdrf5r300}, \ref{gwT75epsa}, and 
\ref{gwr80T75eps} illustrate the effect of a varying dielectric
constant on a CPBG. In the diamond case, the shell dielectric
constant is varied. In the zinc-blende case, one of the spheres 
in the lattice primitive cell remains a silver 
core - $n_s=1.45$ (silica) shell sphere
with fixed $r_c/r_s=0.75$ ($f_m=7.2\%$),
and we vary the dielectric constant of the dielectric sphere.
Quite amazingly, small metal inclusions
have the biggest effect for $\varepsilon\in [2,12]$
(cf. Figs.  \ref{gwT75epsa} and \ref{gwr80T75eps}), 
which is a typical dielectric constant in the near infrared
and in the visible for many materials such as semiconductors and polymers.
The dependence of the CPBG's on the dielectric constant is very complex. 
Figures \ref{gwdmdrf5r300}, \ref{gwT75epsa}, and \ref{gwr80T75eps}
show that the  $2$nd-$3$rd CPBG remains the largest CPBG for 
$\varepsilon\leq 9$.
According to Figs. \ref{gwT75epsa} and \ref{gwr80T75eps}, for zinc-blende 
structures and  $\varepsilon\geq 9$,  the  $2$nd-$3$rd CPBG becomes smaller
than the $5$th-$6$th CPBG. The $8$th-$9$th CPBG remains small for both
diamond and zinc-blende structures. As already seen for a diamond structure
of non-overlapping dielectric spheres (see Fig. \ref{rgwdmd}) \cite{AMS}, 
if $\varepsilon$ increases beyond a certain 
threshold value, the  $2$nd-$3$rd CPBG begins to narrow and may 
even disappear.  Such a narrowing of the  $2$nd-$3$rd CPBG with an increasing 
dielectric constant seems a characteristic feature of the diamond and
zinc-blende structures.  Figure \ref{gwr80T75eps} shows the same kind of 
``gap competition" which has been first observed in the purely 
dielectric case  (see Fig. \ref{rgwdmd}): a CPBG first increases
with increasing dielectric constant, 
but, at some dielectric contrast shortly before another 
CPBG starts to open, the CPBG begins to decrease and it may eventually
disappear.

\subsection{Scalinglike behavior}
\label{ssec:sca}
A further remarkable feature of the metallo-dielectric structure
is a surprising scalinglike behavior,
which is intrinsic only to ideal dispersionless structures.
Given the metal filling fraction $f_m$, the best scalinglike 
behavior is observed in our zinc-blende structures 
(see Figs. \ref{gwT75r},
\ref{midgapsiocpT75r}, \ref{midgapT75}, \ref{dmdsiocpT75RTA}).
A scalinglike behavior is expected once the metal
$\varepsilon$ at a given midgap wavelength becomes sufficiently 
large and negative. Indeed, on the single-scatterer level 
it makes rather little difference if the real part of the
metal $\varepsilon=-200$ or $\varepsilon=-\infty$, 
the limit of a perfect metal. Figure \ref{ssextT75sior} shows that
differences in the sphere extinction efficiency rapidly
decrease with an increasing sphere radius. There is only a little
difference in the extinction efficiencies of $r_s=300$ nm and 
$r_s=600$ nm spheres. One can, therefore, view the preceding figures
involving $r_s=300$ spheres as a limiting case of a perfect metal.
For silver inclusions, the midgap position 
and the $2$nd-$3$rd  CPBG edges follow almost perfect scaling 
down to $r_s=200$ nm.  For $r_s<200$ nm,
the gap width and gap edges begin to depend stronger and stronger
on decreasing $r_s$. However, if the midgap wavelength is plotted in nm 
against sphere radius, one would see a linear dependence down to 
$r_s=80$ nm (see Figs. \ref{midgapT75}, \ref{midgapsio}),
for which the $2$nd-$3$rd CPBG midgap
wavelength is typically below $600$ nm and the real part of silver 
$\varepsilon\approx -10$.
Therefore, in Fig. \ref{midgapsiocpT75r} the ratios of the
midgap and gap edges of the $2$nd-$3$rd CPBG to the sphere radius 
are plotted to make deviations from the scalinglike behavior visible at all.
This scalinglike property is very useful from a practical point of view.
It means that once a CPBG is found, with some midgap wavelength 
$\lambda_c$, the CPBG can be centered at any other wavelength by 
a simple scaling of all the sizes of a structure.

\subsection{Reflectance, transmittance, absorptance}
\label{ssec:rta}
In order to further investigate optical properties
of our photonic structures, reflectance, transmittance, 
and absorptance were calculated. By reflectance, ${\cal R}$, and 
transmittance, ${\cal T}$, we mean total reflected and 
transmitted field (the specular and all propagating higher 
diffraction orders),  respectively. Absorptance, ${\cal A}$, 
is then defined as the total loss
in the structure, ${\cal A}=1-{\cal R}-{\cal T}$.
Reflectance, transmittance, and 
absorptance were calculated by adapting available computer code
\cite{YSM}, which is based on the photonic LKKR 
method \cite{Mod}. The same method was also used in Refs. 
\cite{WCZ} and \cite{ZLW}. To reach convergence within $1\%$ around the 
$2$nd-$3$rd  CPBG, $l_{max}=6$ was used.

For a metallo-dielectric structure is characteristic
that, beyond a certain crystal thickness, reflectance, 
transmittance, and absorptance within a CPBG change 
only negligibly with an increasing crystal thickness \cite{MTC}.
As a function of the crystal thickness, $\ell$, absorptance
behaves as
\begin{equation}
{\cal A}(\ell) = {\cal A}_{sat}[1-\exp(-\ell/\ell_A)],
\end{equation}
where $\ell_A$ is an ``absorption" length and
${\cal A}_{sat}$ is the saturated absorptance. For 
some CPBG's, ${\cal A}_{sat}$ can only be a few per cent.
This saturated behavior can easily be explained as follows.
Because of a CPBG, light can only penetrate the first few layers
of a crystal. Hence, an addition of further layers beyond
the penetration length has only minor influence.
As a direct consequence of the saturation, absorptance
per crystal thickness decreases approximately linearly
with the crystal thickness, 
${\cal A}(\ell)/\ell \approx {\cal A}_{sat}/\ell$. 
For a high metal-volume
fraction, as for an fcc photonic crystal, 
such a saturated behavior can already be established
after three crystal planes \cite{WCZ,ZLW}. In our case, the metal
volume fraction is almost 20 times smaller.
Consequently, the saturation (with two digit precision) 
is established
after 12 crystal planes for a normal incidence
in the (111) crystal direction, or
after two unit-cell thickness. The (111) crystal 
direction has been chosen because this choice makes it 
possible to compare the properties of our photonic crystals 
against those of fcc photonic crystals, which properties
have been studied in the same crystal direction \cite{WCZ,ZLW}.
It is useful to remind 
that the (111) direction corresponds to the 
$\Gamma$-L direction in the notation of special points 
on the surface of the Brillouin zone. Hence, for our diamond
and zinc-blende structures, reflection and transmission measurements 
in the (111) direction probe the slice of band structure (see 
Figs. \ref{diamcosta} and \ref{dmdeps12r80T75}) which consists 
of the line parallel to the $y$ axis which intersects 
the $x$ axis at the L point.

In Fig. \ref{dmdsiocpT75RTA} reflectance, 
transmittance, and absorptance of light incident normally on a 
two unit cells (12 planes) thick zinc-blende lattice of 
spheres in air stacked in the (111) direction is shown.
Note first a scalinglike behavior of 
reflectance, transmittance, and absorptance in 
Fig. \ref{dmdsiocpT75RTA}. 
The spheres are as in Figs. \ref{gwT75r}, \ref{midgapsiocpT75r}, 
\ref{midgapT75}. The AABBCC stacking sequence is depicted in 
Fig. \ref{dmd111stack}. The gap width to midgap frequency 
ratio of the $2$nd-$3$rd CPBG of these configurations has been 
shown in Fig. \ref{gwT75r}. The midgap wavelength of the 
$2$nd-$3$rd CPBG has been shown in Figs. \ref{midgapsiocpT75r}, 
\ref{midgapT75}. 

Calculations based on the LKKR method \cite{YSM,Mod} showed that 
the saturated absorptance within a CPBG can be kept at 
very small levels. Given the desired gap width of $5\%$, 
the smallest absorption was found for close-packed zinc 
blende structures, even though the required $f_m$ was 
typically twice as large as that of the close-packed 
diamond structure.  
According to Fig. \ref{dmdsiocpT75RTA}, the saturated absorptance
within a CPBG of $\geq 5\%$ can be kept below  $5\%$ for a CPBG
centered at $\lambda\approx 600$ nm and is less than $2.6\%$ 
for a CPBG centered at $\lambda\geq 750$ nm. 
This should be tolerable in most practical applications. 
The results on the absorption are by far the best which
have been demonstrated for a 3D metallo-dielectric structure
with a CPBG. They are an order of magnitude better than for 
the case of an fcc lattice of metal coated spheres \cite{WCZ}.
(Although we have used the same method as in Ref. \cite{WCZ,ZLW}
to calculate absorption, when recalculating the results in 
Ref. \cite{WCZ} for an fcc lattice of metal-coated dielectric 
spheres, a much larger absorptance was obtained.)
Our results for the saturated absorptance compare well 
to the best results for one-dimensional (1D) and two-dimensional (2D)
metallo-dielectric photonic crystals which show a saturated absorptance 
of $\approx 1\%$ and $\approx 3\%$, respectively, within a 
CPBG centered at $\lambda\approx
600$ nm \cite{AMs}.  

Note in passing that within the $2$nd-$3$rd CPBG and for normal
incidence in the (111) direction, almost 99\% reflection goes into 
the specular beam. As the angle of incidence increases,
the specular reflectivity decreases and gradually more light
is reflected in nonspecular directions.

\section{Discussion}
\label{sec:disc}

\subsection{Use of others metals as a sphere core}
\label{ssec:other}
So far, when presenting our results on the effect of 
small inclusions of a low absorbing metal on the photonic 
band structure of diamond and zinc-blende photonic crystals 
we have used exclusively silver. This noble metal guarantees 
the best properties of metallo-dielectric photonic crystals 
in the near infrared and in the visible \cite{WCZ,AM2,ZLW}. 
However, almost identical results have also been obtained 
for aluminum, copper, and gold cores, provided that sphere
radius $r_s\geq 250$ nm  (see Fig. \ref{dmdsiocprT75RTA250}). 
For $r_s< 250$ nm spheres, the results for different metal 
can be increasingly different with 
decreasing $r_s$, nevertheless, qualitative features remain 
the same (see Fig. \ref{dmdsiocprT75RTA100}). 

This behavior is explained as follows.
When the scalinglike behavior was discussed in Sec. \ref{ssec:sca},
it was argued that one can view the results
involving $r_s\geq 250$ nm spheres as 
those corresponding to a limiting case of a perfect metal,
when metal dielectric constant is $\varepsilon=-\infty$.
Indeed, Fig. \ref{ssextT75sior} showed that
differences in the single-sphere extinction efficiency rapidly
decrease with an increasing sphere radius. Therefore,
one expects only minor differences for diamond and zinc-blende
photonic crystals with $r_s\geq 250$ nm spheres when gold, copper,
or some other metal is used instead of silver, with the other
lattice and sphere parameters being identical. These expectations
are fully confirmed by exact calculations, and the results are
displayed in Figs. \ref{dmdsiocprT75RTA250}, \ref{dmdsiocprT75RTA100},
A similar behavior, i.e., only a slight change in
absorption when gold or copper has been used
in place of silver, has also been found for an fcc structure\cite{KSB}.

\subsection{Fcc vs diamond and zinc-blende structures}
\label{ssec:fccvdmd}
Obviously, small metal inclusions do not open a CPBG 
in every dielectric structure. For example, a simple fcc
lattice of dielectric spheres with a metal core requires 
$f_m\approx 50\%$ to open a CPBG \cite{AM2,ZLW}. 
For an fcc lattice of metal-coated dielectric 
spheres, the required  metal filling fraction $f_m$ is slightly lower 
but still very high ($\approx 40\%$
\cite{WCZ}) and still almost 40 times larger than in our  
case. Therefore, not surprisingly,
when going further to shorter and shorter wavelengths, one is facing
an increasing absorption: at $\lambda\approx 600$ nm the absorption
exceeds $30\%$ even within a CPBG. 
Although such a metallo-dielectric fcc structure could provide a 
CPBG \cite{WCZ,AM2,ZLW} at near infrared, the extension to the visible 
is difficult. 
A probable reason of such a big effect of small metallic inclusions
on the photonic band gaps of the diamond and 
zinc-blende structures compared to an fcc
structure is because the former have  much better 
photonic band-gap properties already in the limiting cases
of purely dielectric and purely metallic spheres. 
In the limit of dielectric spheres in air \cite{HCS,SBMc}, the threshold 
value of 
the dielectric contrast to open the  $2$nd-$3$rd CPBG for a diamond 
structure is only $5.2$ \cite{AMS}, whereas an fcc structure
does not have any CPBG, irrespective of the sphere dielectric 
constant \cite{MS,SHI}.
In the opposite limit of pure metallic spheres,
the $2$nd-$3$rd CPBG for the diamond lattice
is huge. For silver, depending on the sphere radius $r_s$,
it can stretch from  $60\%$ ($r_s=80$ nm) to $75\%$ 
($r_s\geq 300$ nm) \cite{AMS,ZLW}.
This is consistent with a previous estimate of $g_w\geq 60\%$
for the case of an ideal metal ($\varepsilon_s=-\infty$) \cite{FVJ}.
On the other hand, for an fcc lattice a higher CPBG opens  (between
the $5$th-$6$th bands) and is smaller ($\approx 40\%$) \cite{ZLW}.

\subsection{Behavior of the $2$nd-$3$rd CPBG with 
an increasing metal-volume fraction}
\label{ssec:behxx}
In order to explain the opening of a CPBG with an increasing metal 
volume fraction, the following argument has been used 
by Zhang {\em et al.}\cite{ZLW}: Low-frequency waves propagate through a 3D 
metallo-dielectric structure as in an effective
medium characterized by the effective dielectric constant 
$\varepsilon_{e\!f\!f}$ \cite{MG}. This low-frequency pass band
will have zero group velocity when the wave vector approaches the
Brillouin-zone boundary where the wave vector $k\approx \pi/a$,
$a$ being the lattice constant. Thus the highest frequency $\omega_l$
of this low-frequency pass band 
$\approx c/(a\sqrt{\varepsilon_{e\!f\!f}})$.
On the other hand, metal cores inscribe a void with a scale of the
same order as $a$, and the fundamental resonance modes in the void
should have frequencies of the order $\omega_u\approx c/a$. 
These voids are
always connected together by channels, so that the modes can hop
from one void to another forming a pass band. Since 
$\varepsilon_{e\!f\!f}$
increases when the metal-volume fraction increases, it may be
possible to have $\omega_l<\omega_u$ when the metal-volume fraction
exceeds a certain threshold value. Then, assuming that there
are no other bands between the low-frequency pass 
and hopping bands, a CPBG should be formed 
between $\omega_l$ and $\omega_u$. However, Zhang {\em et al.} \cite{ZLW} 
applied their argument to an fcc lattice where a CPBG opens between 
the $5$th-$6$th bands, i.e., the band on the lower edge of the CPBG 
is not a low frequency pass band. It seems more natural to reconsider
their argument for our case of diamond and zinc-blende photonic
crystals of coated spheres.

The Garnett formula\cite{MG} [Eq. (\ref{maxgsp})] 
yields $\varepsilon_{e\!f\!f}$ for a cubic lattice of
both homogeneous and coated spheres. One has only to substitute
 a corresponding polarization factor for $\alpha$. 
In the case of a coated sphere with a single coating, embedded in 
the host medium characterized by the dielectric constant 
$\varepsilon_h$, let us denote the dielectric constant and radius 
of the sphere  core (shell) by $\varepsilon_c$ and $r_c$
($\varepsilon_s$ and $r_s$), respectively. 
Defining $x=r_c^3/r_s^3$,  
$\alpha_c=(\varepsilon_c-\varepsilon_s)(\varepsilon_c+2\varepsilon_s)$,
and
$\alpha_s=(\varepsilon_s-\varepsilon_h)(\varepsilon_s+2\varepsilon_h)$,
the polarization factor $\alpha$ of a coated sphere with
a single coating can be written as \cite{MS}
\begin{equation}
\alpha =
\frac{\alpha_s+ x\alpha_c(\varepsilon_h+2\varepsilon_s)/
(\varepsilon_s+2\varepsilon_h)}
{1+2x\alpha_c\alpha_s}\cdot
\label{coated}
\end{equation}
Here  $0\leq x\leq 1$ and $0\leq \alpha_j<1$. Note that 
$\alpha_c$  ($\alpha_s$) coincides with the 
polarization factor of a homogeneous sphere
with the dielectric constant $\varepsilon_c$ ($\varepsilon_s$)
embedded in the host medium characterized by the 
dielectric constant $\varepsilon_s$ ($\varepsilon_h$) \cite{MG}.
(Polarization factors of a homogeneous sphere are also recovered 
from $\alpha$ in one 
of the  following limits: $x\rightarrow 0$, $x\rightarrow 1$, 
and $\varepsilon_c\rightarrow\varepsilon_s$ \cite{MS}).

For a diamond lattice of close-packed metal-core  dielectric-shell 
spheres in air, $\varepsilon_{e\!f\!f}$, as calculated from Eqs. 
(\ref{maxgsp}) and (\ref{coated}), increases from $\approx 1.3$ up 
to $\approx 1.9$, i.e., by only a factor of $\approx 1.5$, 
when $r_c/r_s$ increases from $0.1$ to $0.75$ (irrespective of 
the real part of the dielectric constant of the metal core when varied 
between $-50$ and $-500$, since the metal-volume fraction
is kept rather small). Therefore,  in our case, the change in
$\varepsilon_{e\!f\!f}$ with an increasing metal-volume fraction 
$f_m$ is by no means dramatic. Yet the behavior of the $2$nd-$3$rd CPBG
edges, as shown in Fig. \ref{edgesxx}, seems to be qualitatively
in agreement with the argument by Zhang {\em et al.} \cite{ZLW}.
For $r_s=80$ nm spheres, both CPBG edges first decrease with 
increasing $r_c/r_s$. However, when $r_c/r_s$ increases beyond 
$\approx 0.5$, the upper $2$nd-$3$rd CPBG edge slows down
its decrease and becomes approximately 
constant, whereas  the lower edge of the $2$nd-$3$rd CPBG continues 
in its monotonic decrease. The case of $r_s=300$ nm spheres
shows a qualitatively different behavior of the upper $2$nd-$3$rd 
CPBG edge, which frequency increases with increasing $r_c/r_s$.
This behavior is exactly what one would expect according to the argument
by Zhang {\em et al.} \cite{ZLW}. 
In the case of $r_s=300$ nm spheres, the $2$nd-$3$rd
CPBG is located at the frequency range for which silver skin depth
[$=c/(2\omega\, \mbox{Im}\, n(\omega))$]
is only slightly above $10$ nm, whereas the silver core has radius
between $150$-$180$ nm. Therefore, light can only penetrate
a small fraction of the silver core. Consequently, since the void 
volume between spheres shrinks as $r_c/r_s$ increases, the fundamental 
resonance modes in the void shift to higher frequencies.
In the case of $r_s=80$ nm spheres, the $2$nd-$3$rd CPBG is 
located at the frequency range for which silver skin depth 
can be as large as $50$ nm, i.e., 
light can penetrate entire silver core. Therefore, the ``void"
part of the argument by Zhang {\em et al.} is much less applicable 
in the $r_s=80$ nm case.

\subsection{Coated versus homogeneous metal spheres}
\label{ssec:cvhom}
We have also examined if a large and robust CPBG can open by using 
homogeneous metal nanospheres arranged on a diamond lattice 
in a dielectric, with the possibility of fabricating a diamond
structure by layer-by-layer deposition. According to 
Fig. \ref{dmd111stack}, the sphere projections on the stacking
$z$ direction in neighboring planes do not overlap if the sphere
radius $r_s\leq r_{cp}/3$, where $r_{cp}$ is the sphere radius
in the close-packed case. Therefore, if $r_s\leq r_{cp}/3$ 
it would be possible to fabricate a diamond lattice by 
depositing first a hexagonal plane of spheres on a substrate 
and then filling in the interstitial by a dielectric untill all 
spheres are embedded in a dielectric layer. One would 
then repeat the procedure layer by layer and grow gradually a 
diamond lattice. Unfortunately, the constraint $r_s\leq r_{cp}/3$ 
means that the resulting sphere filling fraction $f_s\leq 1.26\%$, 
which is too small to achieve a large and robust CPBG with 
homogeneous metal nanospheres in a dielectric.
Indeed, no CPBG was found for a diamond structure of 
homogeneous metal nanospheres with $r_s\leq 0.45 r_{cp}$ ($f_m\approx 3.1\%$)
embedded in a dielectric  host with the dielectric constant 
$\varepsilon=2.1$, $3$, $4$, and $6.25$. Photonic 
band-structure calculations revealed that one needs 
$f_s=f_m\approx 10\%$ to open the $2$nd-$3$rd CPBG of $5\%$ 
with metal nanospheres in silica. For comparison, note that 
for a close-packed diamond structure of $n_s=1.45$ dielectric 
spheres with a metal core and radius of $80$ nm, the $2$nd-$3$rd CPBG 
opens for $r_c=0.32 r_{cp}$, and for $r_c=0.45 r_{cp}$ 
the $2$nd-$3$rd CPBG reaches almost $5\%$  (Fig. \ref{gwAgr80xx}). 
This suggests that a metal 
core-dielectric shell sphere morphology  plays a crucial
role in the formation of CPBG's. Indeed, if the same volume 
of a metal is spread homogeneously within the spheres, no 
CPBG opens in the spectrum. In the latter case,
the sphere dielectric constant was calculated using the Garnett 
formula (\ref{maxgsp})  \cite{MG}.

A dielectric contrast between the dielectric shell and the  host 
dielectric medium also seems to be  an important factor.
Figure \ref{gwepsc8eps} shows the dependence of the $2$nd-$3$rd CPBG 
for a close-packed metallo-dielectric diamond lattice of spheres 
of radii $r_s=80$ nm and $300$ nm
on the host dielectric constant. Spheres have a silver core and 
dielectric shell with $\varepsilon=8$, with the core to 
total sphere radii ratio either $r_c/r_s=0.5$ ($f_m=4.25\%$), or
$r_c/r_s=0.6$ ($f_m=7.3\%$). 
The shell dielectric constant has to be
sufficiently large to open a CPBG. For example, for 
the dielectric 
shell with $\varepsilon=4$ and $r_c/r_s=0.4$ ($f_m=2.2\%$), 
the $2$nd-$3$rd CPBG is closed in the range of the host 
dielectric constant in Fig. \ref{gwepsc8eps}.

\subsection{Absorption}
\label{ssec:abs}
Similarly, as in the case of a photonic crystal 
of small $r_s=5$ nm Drude plasma spheres \cite{YMS}, one observes 
(see Figs. \ref{dmdsiocpT75RTA}, \ref{dmdsiocprT75RTA250}, 
\ref{dmdsiocprT75RTA100}) that the absorptance peaks coincide 
with the transmittance peaks. An explanation for this behavior can 
already be traced down to the case of a homogeneous slab
of metal. In the latter case, one can show that neither the imaginary 
part of the metal dielectric constant nor the inverse skin depth 
(absorption coefficient) characterize properly the absorption of 
an incident light. Indeed, the absorption coefficient is a measure of
absorption of only that light which has entered the metal.
Higher transmittance means higher probability of light being
absorbed. Hence, a proper measure of the degree of absorption 
of an incident light is rather a product of the transmittance 
and the absorption coefficient. 

In the purely dielectric case \cite{YSM} and in the case of an
fcc crystal of small $r_s=5$ nm Drude plasma spheres with 
sufficiently large relaxation time of conduction electrons 
and a low volume fraction ($f=0.1$) \cite{YMS}, transmittance 
peaks occur when the effective half wavelength,
$\lambda/2=\pi/k_\perp$, $k_\perp$ being the wave vector
in the stacking direction, times an integer equals approximately 
the thickness of the photonic crystal slab. Hence,
when transmittance is plotted against $k_\perp$, one observes
a series of almost equidistantly spaced peaks \cite{YSM,YMS}.
It is well known that $dk_\perp/d\omega$ can attain 
very high values near band edges. As a rule of thumb,
band edges are located either at the surface, 
or in the center (${\bf k}=0$ or $\Gamma$ point) 
of the Brillouin zone.  At the points on the surface of the 
Brillouin zone with reflection symmetry, frequency eigenvalues
satisfy $\omega({\bf k})=\omega(-{\bf k})$. For those points
$dk_\perp/d\omega$ even diverges to infinity. (The latter is
equivalent to saying that at such a band edge the group velocity of
light approaches zero.) Therefore, if transmittance is plotted against 
frequency, one observes a considerable increase in the density 
of transmittance peaks in the immediate vicinity of 
a frequency gap. However, when the relaxation time of conduction 
electrons of Drude plasma spheres gradually decreases (what is
equivalent to an increasing imaginary part of the dielectric
constant), the densely spaced sharp transmittance and absorptance peaks 
near band edges merge and generate a rather smooth profile \cite{YMS}.
This is what is also observed in our case, as
exemplified in Figs. \ref{dmdsiocpT75RTA}, \ref{dmdsiocprT75RTA250}, 
and \ref{dmdsiocprT75RTA100}, which show no densely spaced and
sharp transmission peaks in the proximity of the lower 
$2$nd-$3$rd CPBG edge. 

A dramatic change in the absorptance at the lower and upper 
band gap edges is a reminiscent of the Borrmann effect, previously 
observed in x-ray scattering \cite{Zach}. The origin of this 
effect is that fields in the proximity of band-gap edges 
approach a standing  wave. In the proximity of one band-gap edge, 
fields are mostly localized in 
the regions of a low dielectric constant, 
whereas in the proximity of the other band-gap edge, fields are 
mostly localized in the regions of a high dielectric constant.
This sort of behavior can already be observed for a simple
1D periodic Bragg stack of dielectric layers \cite{Ru}.
Figures \ref{dmdsiocpT75RTA}, \ref{dmdsiocprT75RTA250}, and 
\ref{dmdsiocprT75RTA100} suggest that in the proximity
of the lower band gap edge of our zinc-blende
structures electric field is mostly localized in air,
whereas in the proximity of the upper band gap edge  electric 
field is mostly localized at sphere positions, leading to
a significantly higher absorption.

Our results on absorption compare well to the best results
for 1D and 2D structures \cite{AMs}. In these respective cases,
the absorption of an optimized metallo-dielectric structure can be 
$\approx 1\%$ and $\approx 3\%$ within a CPBG
centered at $\lambda\approx 600$ nm. 
Given the metal filling fraction $f_m$, it has been found 
that absorption is reduced when metal forms isolated islands \cite{KSB}.
This is an expected result: when metal forms an interconnected
network, DC conductivity is nonzero and long-range conduction
currents are induced which lead to higher losses.
Our results suggest that absorption is reduced further
when separation of metal islands increases.
Indeed, a larger separation of metal cores in our 
zinc-blende structure compared to a diamond structure  
resulted in a significantly reduced absorption.
This is probably because of a reduction of near-field electromagnetic 
energy transfer between the metal cores with an increased
separation of the metal islands in the structure.

\subsection{Fabrication}
\label{ssec:fabr}
Whereas a simple fcc structure occurs naturally
in colloidal crystals formed by monodisperse colloidal suspensions 
of microspheres  or  via template-directed colloidal crystallization
\cite{BRW}, diamond and zinc-blende structures are very difficult to
fabricate. Nevertheless, several new ways have recently been 
proposed to make these and other complex
structures \cite{AvB,SLM,VCD} and therefore  diamond and zinc-blende 
structures deserve our full attention. One of the possibilities is a
``do-it-yourself organization'' which, using artificial templates,
enables one to create structures beyond thermodynamic limits \cite{AvB}.
Patterning of a surface can be performed using optical tweezers
\cite{HVF} and particle manipulation can also be employed in a 
microrobotic technique \cite{SLM}. Difficulties in optical trapping of a 
metal-core dielectric-shell sphere increase with an increasing core size. 
There is always a danger that a metal core can be melted in the laser focus.
For spheres with a $7.5$ nm gold core and a
silica shell with a final radius of $79$ nm, optical trapping has been
recently  demonstrated experimentally in  Ref. \cite{HVF}.
In Fig. \ref{opttraprf3} we show  that metal-core silica-shell spheres
with much larger cores  can also be optically trapped.
We plot there the axial trapping efficiency $Q$ for
 silver core - $n_s=1.45$ (silica) 
shell spheres with fixed $r_c/r_s=0.3$ immersed in water ($n_h=1.33$).
The axial trapping efficiency $Q$, or the dimensionless normalized axial 
force, when multiplied by the focused laser beam power $P$,
determines the resulting axial force $F$ on a sphere immersed in a 
homogeneous medium with refractive index $n_h$,
\begin{equation}
F=(n_h/c) PQ.
\end{equation}
The laser wavelength $\lambda$ is increased  by $100$ nm with increasing 
sphere radius, from $\lambda=600$ nm for $r_s=80$ nm to 
$\lambda=1000$ nm for $r_s=200$ nm. $Q$ is plotted against $q/r_s$, the
center offset from the focus in units of the sphere radius.
The beam opening angle of the focused beam was taken to be 
$\theta_0=78$ and the ratio of the objective focal length to 
the beam waist was set to $0.99$. The normalized axial force $Q$
was calculated by adapting results of Ref. \cite{MNN}.
A stable equilibrium location on the beam axis corresponds to 
the point where the normalized axial force $Q$, considered 
as a function of $q/r_s$, crosses zero with a negative derivative. 
Fig. \ref{opttraprf3} shows 
that the spheres can be optically trapped for all radii, with a 
stable equilibrium at around $q/r_s\approx 2$.
Note that the equilibrium point is almost five times 
further from the focus in the current case than in the case of purely 
dielectric spheres (cf. Fig. 2 of Ref. \cite{MNN}). A larger
distance from the focus is convenient experimentally as it 
reduces the chance of melting a metal core by an intense light.

\subsection{Related work}
\label{ssec:work}
In our study we have only considered metal-core dielectric-shell
particles which can easily be fabricated \cite{MGM}.
In this respect, our results regarding a CPBG are not optimized.
Probably the required metal-volume fraction can be rendered even
smaller than in our case. One way is to use perturbation theory
along the lines of Ref. \cite{ZZL}. Here it has been demonstrated that 
a photonic band gap can be enlarged by placing metal insertions 
at particular positions of a lattice \cite{ZZL}. However, 
this work dealt with an enlargement of already existing band gaps, 
whereas, in our case, small inclusions of a low absorbing metal 
are used to open a CPBG. Also Ref. \cite{ZZL}
only deals with photonic crystals in two dimensions, whereas 
our work covers 3D photonic crystals. 
Photonic band gaps of AB(3) and B-3 structures of metallo-dielectric spheres
have been investigated in Ref. \cite{FZW}.
It has been shown that an AB(3) photonic crystal contains a large
photonic band gap and a B-3 photonic crystal has more 
than one gap when the filling ratio of metal spheres 
exceeds a threshold. A complementary case of 3D diamond structures
of  metal-coated dielectric spheres has recently been discussed
in Ref. \cite{ZWHM}. Such metal nanoshells, consisting 
of a dielectric core with a metallic shell of nanometer 
thickness (cca $50$ nm), 
can be used to vary the optical resonances of such nanoparticles 
over hundreds of nanometers in wavelength by varying the relative 
dimensions of the core and shell \cite{OAW}. It 
has been shown that, under certain conditions, it is possible 
to center a resonance band, which originates from the single-sphere Mie
resonances of metal-coated dielectric spheres, at the middle of 
the $2$nd-$3$rd CPBG. 
The position and width of the resonance-induced 
in-gap modes can be controlled by making appropriate
choices of the dielectric constant of the inner dielectric sphere and
thickness of the metal-coating layer, respectively. 
Such properties can be
very useful in making optical band filters as well as 
microcavity lasers if
they are sustained in the presence of dissipation. In the case of 
metal-core dielectric-shell spheres the tunability of the 
single-sphere Mie resonances is much smaller. Nevertheless, in the
latter case the resonance-induced in-gap modes
can be observed for certain diamond and zinc-blende configurations.
A study of these resonance-induced in-gap modes in the case of 
metal-core dielectric-shell spheres will be presented elsewhere.
Note that in order to center a single-scatterer resonance 
band within a CPBG, the presence of a metal is crucial. 
The latter is impossible in a purely dielectric case for a reasonable
dielectric contrast \cite{MT}.

\section{Conclusions}
\label{sec:conc}
Diamond and zinc-blende photonic crystals have been studied 
both in the purely dielectric case and in the presence
of small inclusions of a low absorbing metal.
In the former case we have shown that earlier plane wave
method calculations of the photonic band structure were not converged.
Unlike the case of a simple lattice (one scatterer per lattice 
primitive cell) \cite{MS}, for the case of a diamond lattice of 
dielectric spheres, even when using the plane-wave method based 
MIT {\em ab initio} program \cite{JJ}, one has to take a much
higher number of plane waves than expected to reach a convergence 
comparable with the photonic KKR method. To reach convergence of the
photonic band structure of a diamond lattice of dielectric spheres within
$1\%$ the  number of plane waves  $N_c$ has to exceed  $32\, 768$ 
(cf. Ref.  \cite{HCS}) and still an extrapolation 
$N\rightarrow \infty$ \cite{SHI} has to be performed \cite{Meg}.

It has been shown that small inclusions of a low absorbing metal
can have a dramatic effect on the photonic band structure of diamond and 
zinc-blende structures. Several complete photonic band gaps (CPBG's) 
can open in the spectrum, between the $2$nd-$3$rd, $5$th-$6$th, 
and $8$th-$9$th bands. The respective metal-volume fractions
to open a CPBG and to have a CPBG of 5\% can be more than $40$ times 
and $25$ times smaller than in the case of an fcc lattice 
\cite{WCZ,AM2,ZLW}. Unlike in the purely dielectric case,
in the presence of small inclusions of a low absorbing metal
and for a moderate dielectric constant ($\varepsilon\leq 10$), 
the largest CPBG turns out to be the $2$nd-$3$rd CPBG. 
The $2$nd-$3$rd CPBG is the most important CPBG, because it is 
the most stable against disorder \cite{ZLZ}.
One can create a sizeable ($\geq 5\%$) and robust $2$nd-$3$rd CPBG 
at the near infrared and visible wavelengths even for a 
photonic crystal which 
contains more than 97\% of low dielectric constant materials 
($\varepsilon\leq 2.1$). 
For diamond and zinc-blende structures of 
nonoverlapping dielectric and metallo-dielectric spheres, 
a CPBG begins to decrease with an increasing dielectric contrast 
roughly at the point where another CPBG starts to open -- 
a kind of gap competition. 
A CPBG can even shrink to zero when the dielectric contrast
increases further (Figs. \ref{rgwdmd}, \ref{gwT75epsa}, 
\ref{gwr80T75eps}). As a consequence of the gap competition,
the inclusions of a low absorbing metal have the biggest effect 
for the dielectric constant $\varepsilon\in [2,12]$, which is
a typical dielectric constant at near infrared
and in the visible for many materials, including
semiconductors and polymers.
Absorptance in  the $2$nd-$3$rd CPBG of $5\%$  and located above
$\lambda\geq 750$ nm remains very small ($\leq 2.6\%$)
and is an order of magnitude smaller than in the case of an 
fcc lattice \cite{WCZ,ZLW}. Our zinc-blende structures
require an almost twice as large $f_m$ as diamond structures do to 
achieve a comparable CPBG, yet they display smaller 
absorption due to an increased separation of the metal cores. 
The metallo-dielectric structures display a scalinglike
behavior,  which makes it possible to scale the CPBG from microwaves
down to the ultraviolet  wavelengths. 
Given the metal filling fraction $f_m$, this scalinglike
behavior is strongest for zinc-blende structures.

Our results imply that any dielectric material can be used to 
fabricate a photonic crystal with a sizeable and robust CPBG 
in three dimensions, as long as metal inclusions can be added.
These findings (i) open the door for any semiconductor
and polymer material to be used as a genuine 
building block for the creation of photonic crystals with a CPBG 
and (ii) significantly increase the possibilities
for experimentalists to realize a sizeable and robust CPBG at
near infrared and in the visible. By relaxing the constraints
on the dielectric constant, the ultimate goal of matching the
photonic and electronic band gaps, which is a prerequisite
for many applications \cite{Y}, can be achieved.

\section{Acknowledgments}
I would like to thank my colleagues A. van Blaaderen, 
A. Imhof, M. Megens, A. Tip, and K. P. Velikov for 
careful reading of the  manuscript and useful comments. 
SARA computer facilities are also gratefully acknowledged.

\appendix

\section{The KKR method}
\label{app:kkr}
Let $\Lambda$ be a simple (Bravais) periodic lattice.
According to the Bloch theorem, propagating wave $\psi$ in a 
periodic structure with the symmetry $\Lambda$ is characterized by the 
Bloch momentum ${\bf k}$. The latter describes translational 
properties of $\psi$ by any lattice vector ${\bf r}_s\in\Lambda$,
\begin{equation}
\psi({\bf r}+{\bf r}_s)=\psi({\bf r})\exp(i{\bf k}\cdot{\bf r}_s).
\label{blochprop}
\end{equation}
The Bloch property holds irrespective of the spin of a wave, i.e., is the
same for scalar and vector waves.
We shall confine ourselves to the case when, outside scatterers,
wavefunction $\psi$ satisfies the scalar Helmholtz equation, 
\begin{equation} 
[\Delta+\sigma^2]\psi=0, 
\label{scalhlm} 
\end{equation} 
with $\sigma$ being a positive constant. 
Let $G_{0\Lambda}(\sigma,{\bf k},{\bf R})$ denote the free-space periodic 
Green's function of the Helmholtz equation. The latter is defined as
\begin{equation}  
G_{0\Lambda}(\sigma,{\bf k},{\bf R})= 
\sum_{{\bf r}_s\in\Lambda} G_0(\sigma,{\bf R}-{\bf r}_s) e^{i{\bf k}
\cdot{\bf r}_s}  
  =\sum_{{\bf r}_s\in\Lambda} G_0(\sigma,{\bf R}+{\bf r}_s) 
  e^{-i{\bf k}\cdot{\bf r}_s},  
\label{lgrfdf}  
\end{equation}
where  ${\bf R}={\bf r}-{\bf r}'$ and $G_0$ denotes a 
free-space scattering  Green's function of the scalar Helmholtz equation  
at the points ${\bf r}$ and ${\bf r}'$. In 3D
\begin{equation}
G_0(\sigma,{\bf R})=G_0(\sigma,{\bf r},{\bf r}')=
-\frac{\exp(i\sigma R)}{4\pi R}, 
\end{equation} 
where  $R=|{\bf R}|$.
Within the Korringa-Kohn-Rostocker (KKR) method \cite{KR}, band structure 
is determined by solving the KKR secular equation
\begin{equation}
\det \left[1-t(\sigma) g(\sigma,{\bf k})\right]=0,
\label{kkrsec}
\end{equation}
where $t$ is a single-scatterer T matrix and $g$
is the matrix of structure constants \cite{KR}.
Both  $t$ and $g$ in Eq. (\ref{kkrsec}) are considered as matrices 
with matrix elements labeled by pairs of angular momentum numbers 
 $(lm,l'm')$, where $-l\leq m\leq l$.
In the scalar case, for instance, in the case of multiple-scattering
scattering of electrons, the matrix elements of $g$ in the 
angular-momentum basis are defined as expansion coefficients of 
\begin{equation}
G_{0\Lambda}(\sigma,{\bf k},{\bf R}) - G_0(\sigma,{\bf R}) =
 \sum_{lm,l'm'} g_{lm,l'm'}(\sigma,{\bf k}) 
j_{l}(\sigma  r) Y_{lm}({\bf r})j_{l'}(\sigma r') Y_{l'm'}^*({\bf r}'),
\label{dstrco}  
\end{equation}
where $j_l$ are the regular spherical Bessel functions \cite{AS}, and
$Y_{lm}$ are the conventional spherical harmonics \cite{AS}.  

It is interesting to note that, to a large extent, the scalar 
case also covers  the scattering of 
vector and tensorial waves, i.e., waves with a nonzero spin, 
provided each field component  independently obeys 
the scalar Helmholtz equation (\ref{scalhlm}) \cite{Mo,Mod}.
In the vector case of multiple-scattering of electromagnetic waves,
$\sigma=\omega \sqrt{\varepsilon_h}/c$, 
where $\omega$ is the angular frequency, $c$ is the speed of light in 
vacuum, and $\varepsilon_h$ is the dielectric constant of the host
dielectric medium. Let $L=(lm)$ stand
for a multiindex of angular momentum numbers and let $A$ label independent 
polarizations. The corresponding vector structure constants 
$G_{AL,A'L'}$ are obtained from the scalar structure constants 
$g_{L,L'}$ as 
\begin{equation} 
G_{AL,A'L'}= \sum_{p,p'=-1}^1 \sum_{\alpha,\alpha'=-1}^1  
U_A(l,m,p,\alpha) g_{l+p,m+\alpha;l'+p',m'+\alpha'} U_{A'}(l',m',p',\alpha'),  
\end{equation} 
where the $U_A$'s are group-theoretical coefficients, 
in the current case determined by the vector-coupling  
Clebsh-Gordon coefficients\cite{Mo,Mod}. 
Similar for the acoustic and elastic waves. In the case of the
layer KKR method \cite{Mo,YSM}, which can be viewed as
an extension of dynamical x-ray diffraction theory \cite{Zach}
to infinitely many diffracted beams, one is mainly interested
in the reflection, transmission, and absorption properties
of finite-width crystals.  These properties are determined
by the crystal scattering matrix. Roughly speaking, the denominator
of the scattering matrix is the matrix which stands
behind determinant sign in the KKR secular equation (\ref{kkrsec}),
whereas the nominator of the scattering matrix is rather trivial.
Therefore, in a numerical implementation, it is only required 
to make the following two modifications in the scalar KKR and
LKKR numerical codes:
(i) to include a single routine which performs the transformation of 
the scalar  structure constants into vectorial structure constants 
\cite{MS,Mo,AM2} and (ii) to use the routine which calculates $t$, the
single-scatterer T matrix, appropriate to a given boundary
value problem. The same is also true if one
wants to adapt the scalar 3D KKR numerical codes which
deal with clusters of more than 1000 particles of arbitrary shape 
and clusters of more than 1000 arbitrary impurities in a crystal 
\cite{BGZ,F}. There seems to be no obstacle to 
have the photonic KKR methods \cite{MS,Mo,AM2,Mod}
dealing with clusters of more than 300 particles. The KKR 
methods can be used for scatterers of arbitrary shape 
\cite{BGZ} and are optimized for lattices of spheres. 

If $\Lambda$ is not a simple (Bravais)  periodic 
lattice, i.e., there is more than one scatterer in 
the primitive lattice cell, the matrices $t$ and $g$ in
Eq. (\ref{kkrsec}) become matrices with entries labeled by multiindices
$AL \alpha$, where $A$ and $L$ are as before and $\alpha$ runs over all
the scatterers in  the primitive lattice cell \cite{Seg}.

\newpage

\begin{center}
{\large\bf Figure captions}
\end{center}

\vspace*{2cm}

\noindent {\bf Figure 1 -}
Photonic band structure for a close-packed diamond lattice of
spheres of dielectric constant $\varepsilon=12.96$ in air 
(the same set of parameters as in Ref. \cite{HCS}).
Two CPBG open in the spectrum with the
respective gap to midgap frequency ratios of $1.3\%$ and $4.2\%$.
Frequency is in units [c/A], where $c$ is the speed of light in vacuum 
and $A$ is the lattice constant of a conventional cubic unit 
cell of the diamond lattice. If you prefer to use wavelength,
the values on the $y$ axis correspond to those of
$A/\lambda$, where $\lambda$ is the vacuum wavelength.

\noindent {\bf Figure 2 -}
Gap width to midgap frequency ratio for a diamond lattice
of dielectric spheres as a function of the dielectric contrast
$\varepsilon_s/\varepsilon_h$ for sphere filling fractions
$f_s=0.17$ and $f_s=0.34$.

\begin{figure}[tbp]
\begin{center}
\epsfig{file=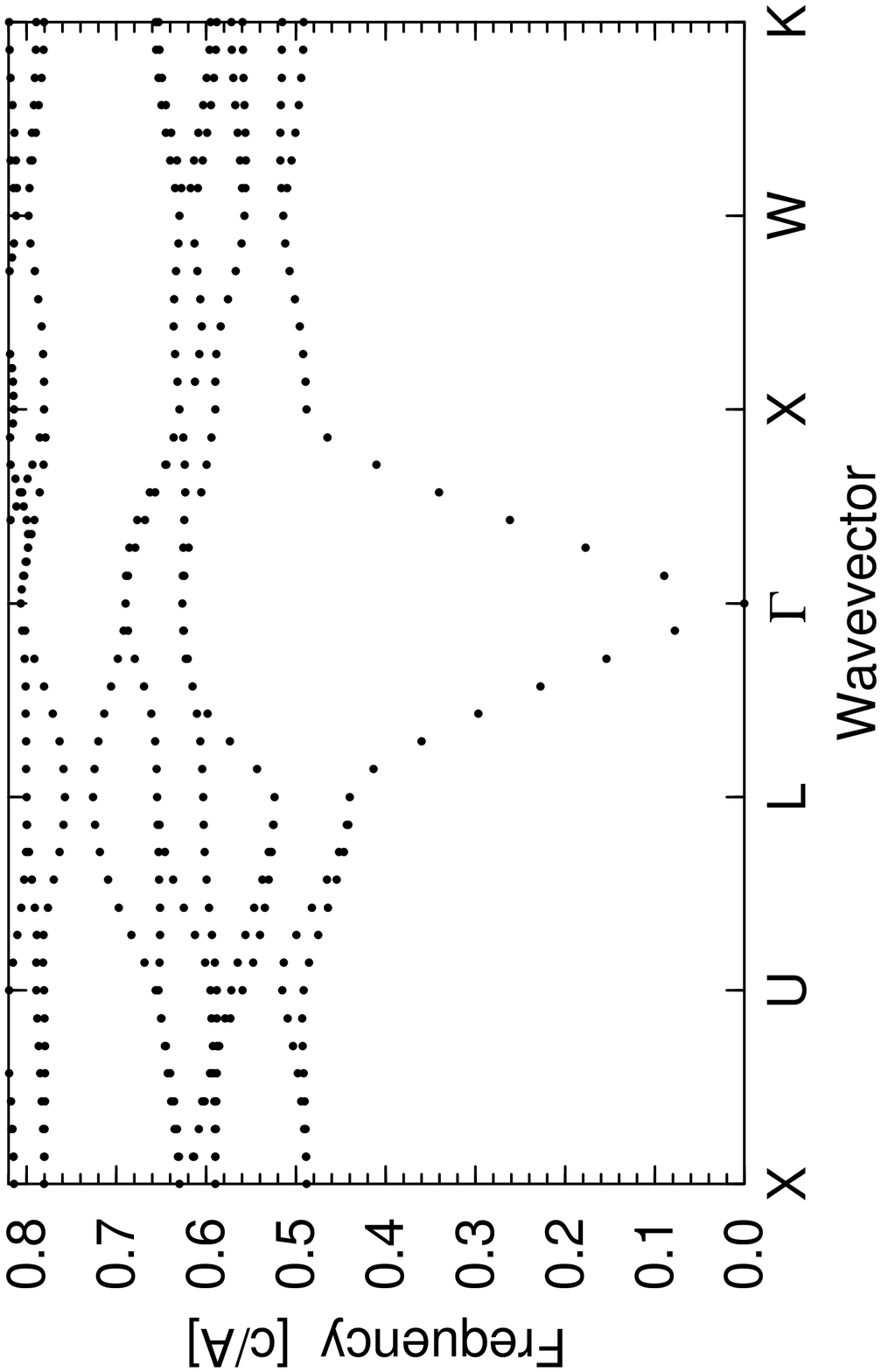,width=12cm,clip=0,angle=0}
\end{center}
\caption{}
\label{diamcosta}
\end{figure}

\begin{figure}[tbp]
\begin{center}
\epsfig{file=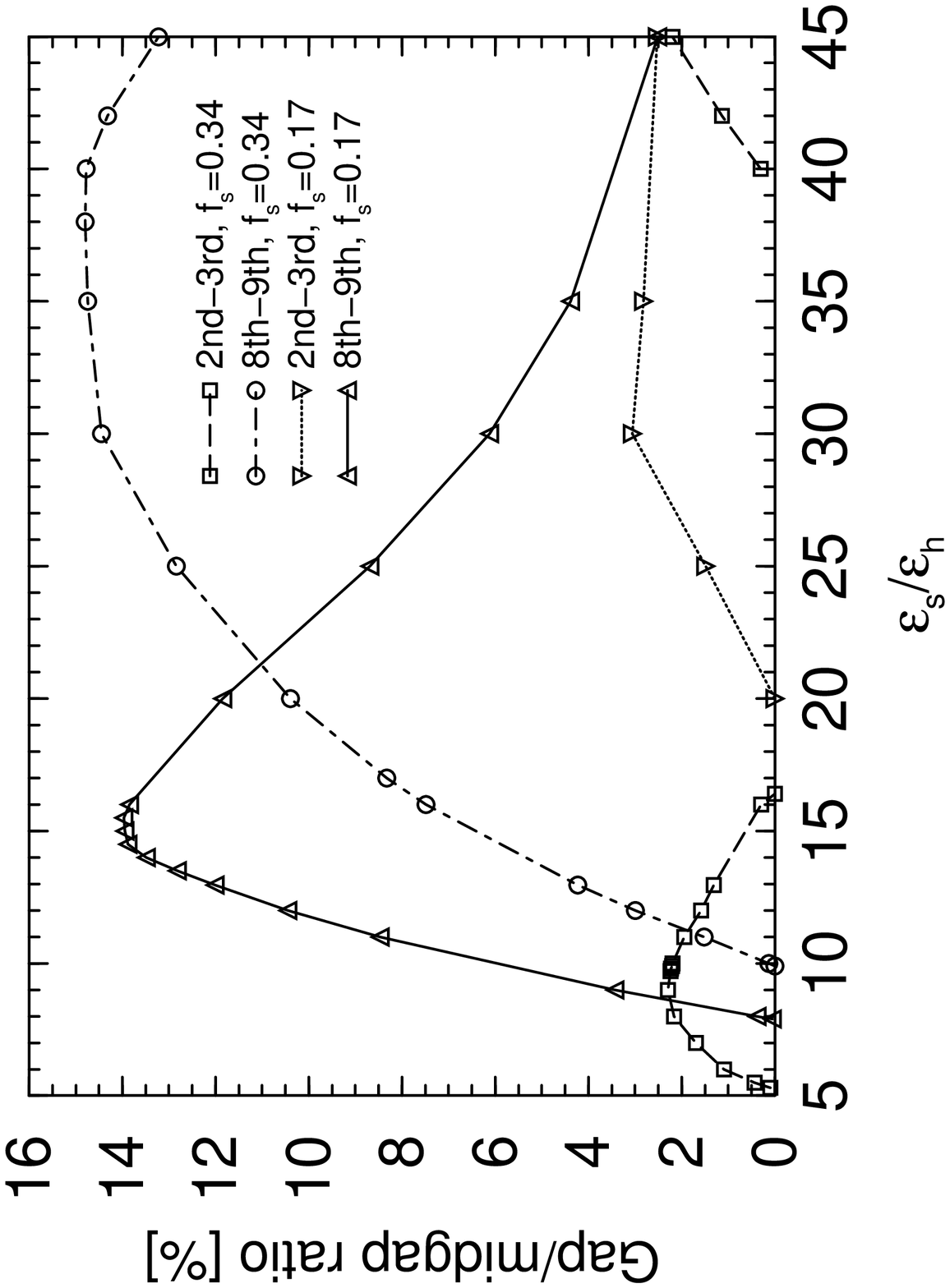,width=12cm,clip=0,angle=0}
\end{center}
\caption{} 
\label{rgwdmd}
\end{figure}

\begin{figure}[tbp]
\begin{center}
\epsfig{file=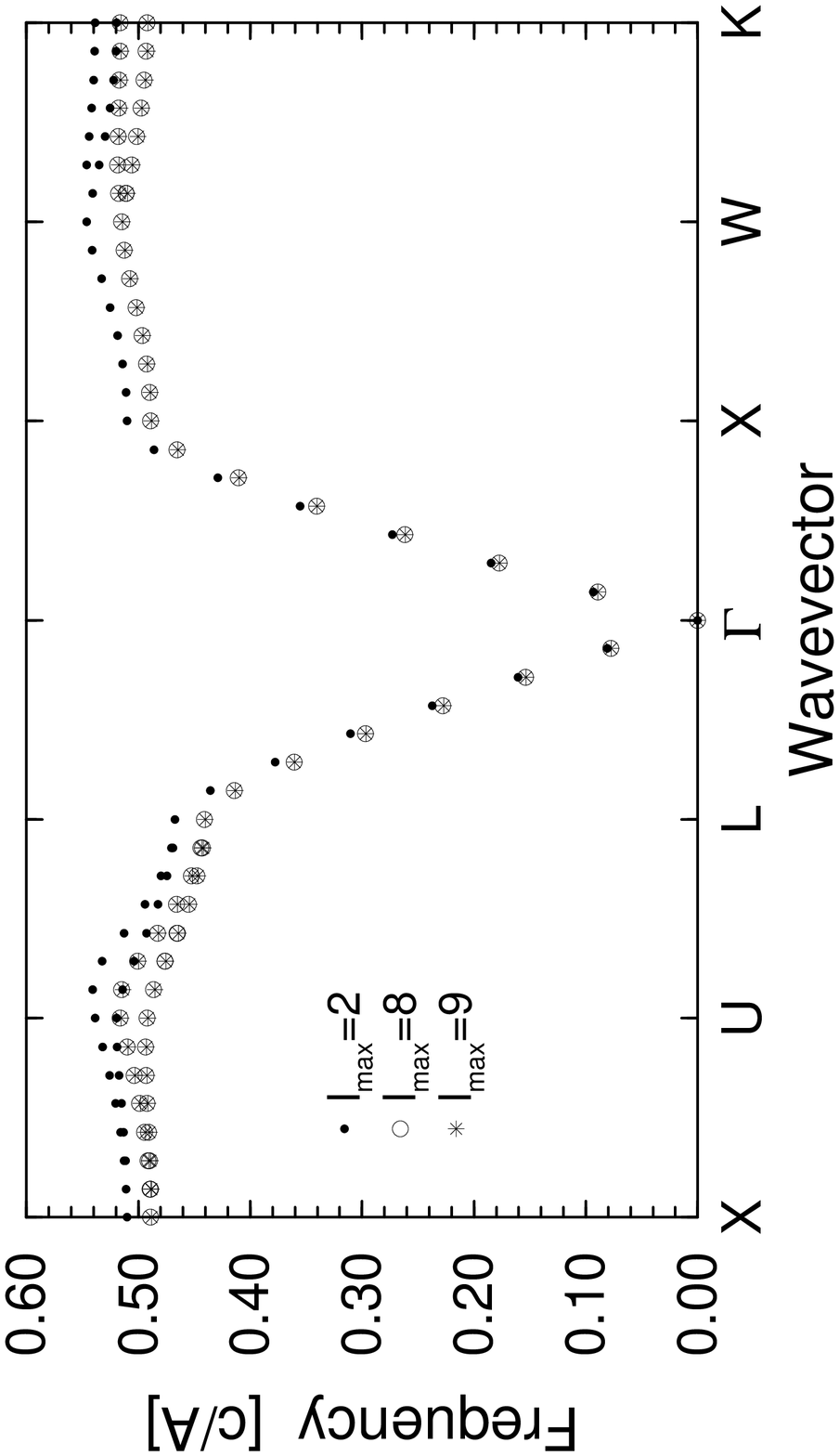,width=12cm,clip=0,angle=0}
\end{center}
\caption{} 
\label{dmdconv}
\end{figure}

\begin{figure}[tbp]
\begin{center}
\epsfig{file=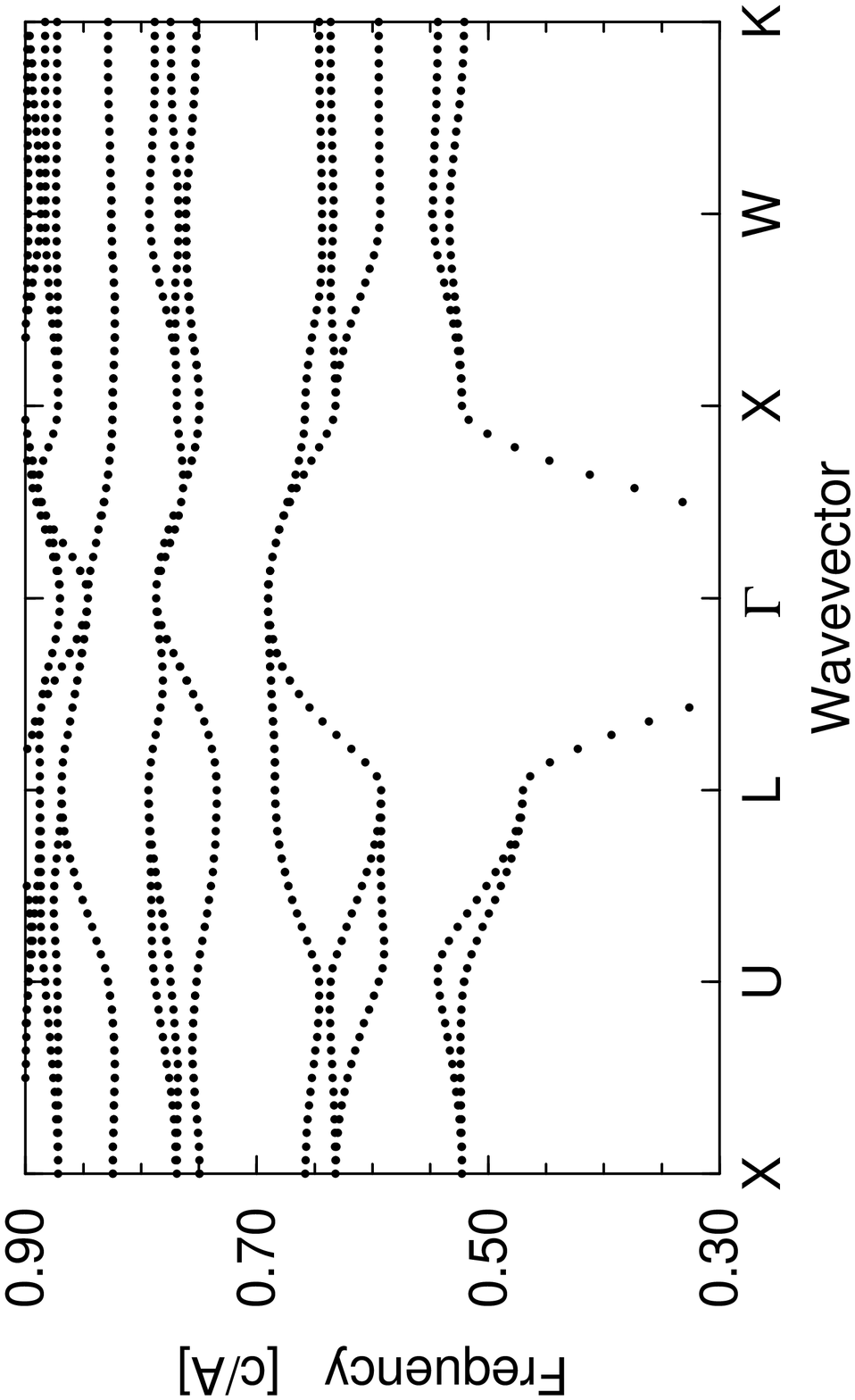,width=12cm,clip=0,angle=0}
\end{center}
\caption{}
\label{dmdeps12r80T75}
\end{figure}

\begin{figure}[tbp]
\begin{center}
\epsfig{file=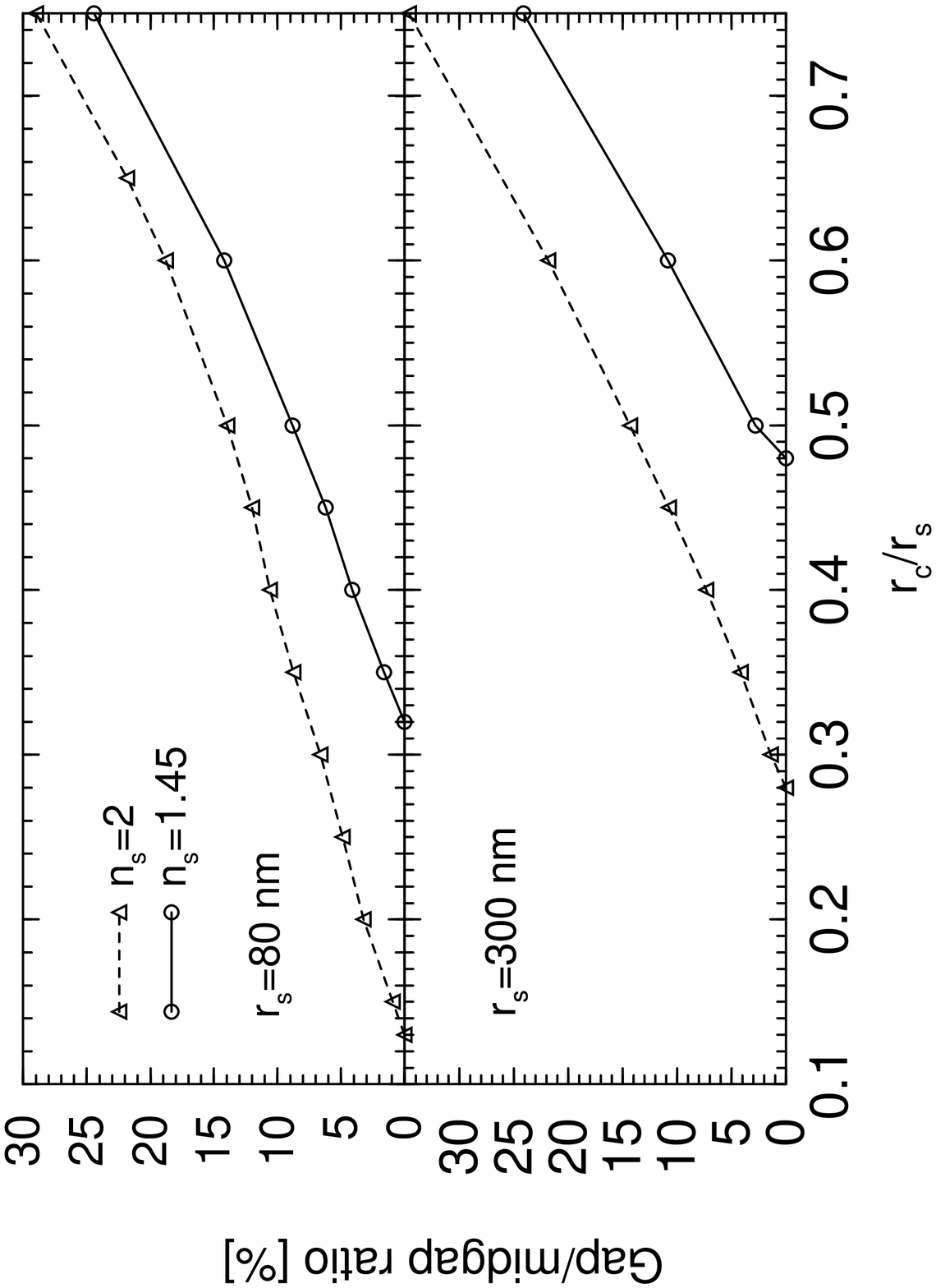,width=12cm,clip=0,angle=0}
\end{center}
\caption{
} 
\label{gwAgr80xx}
\end{figure}

\begin{figure}[tbp]
\begin{center}
\epsfig{file=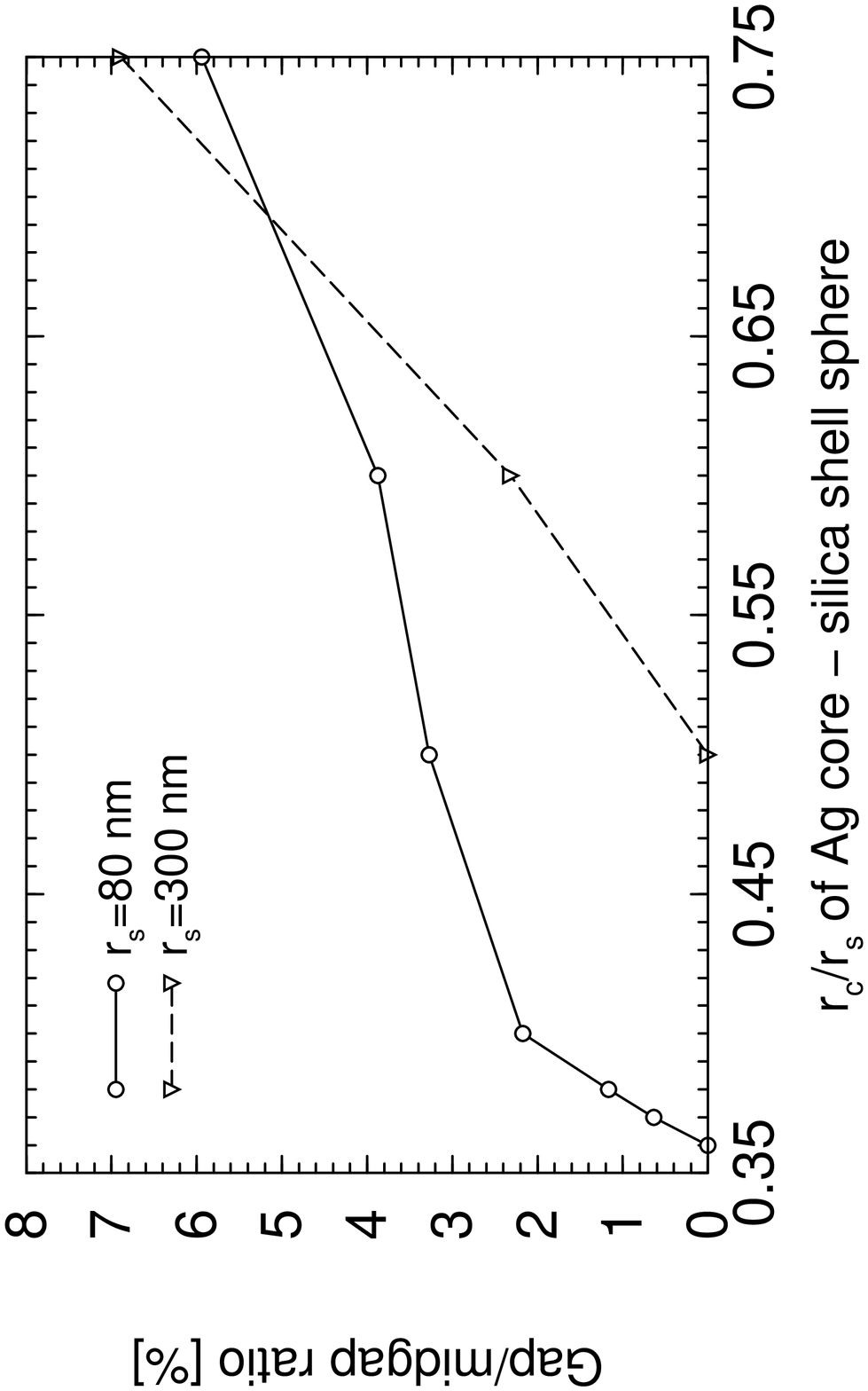,width=12cm,clip=0,angle=0}
\end{center}
\caption{} 
\label{gwTxxsioa}
\end{figure}

\begin{figure}[tbp]
\begin{center}
\epsfig{file=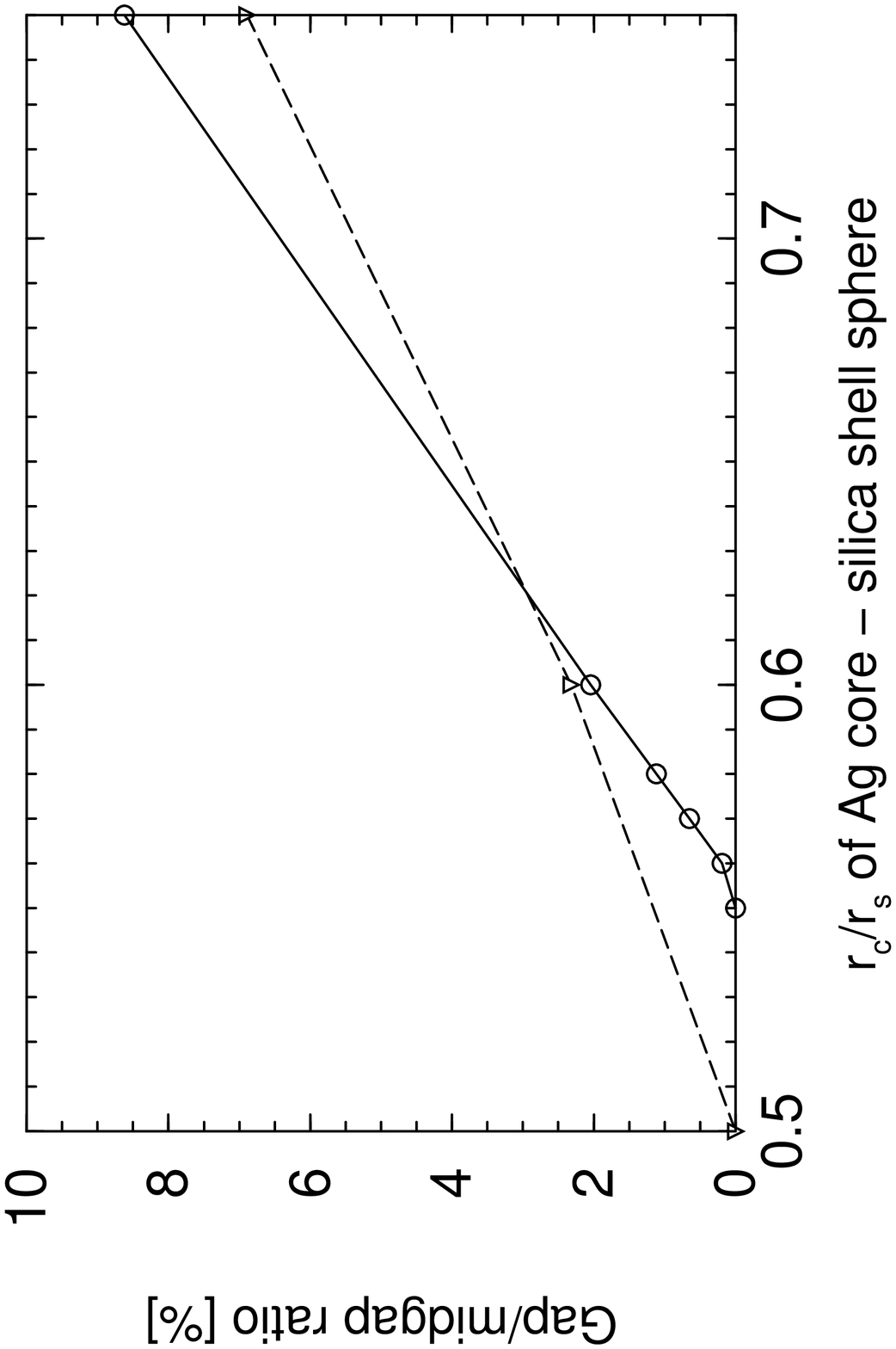,width=12cm,clip=0,angle=0}
\end{center}
\caption{} 
\label{gw300Txx}
\end{figure}

\begin{figure}[tbp]
\begin{center}
\epsfig{file=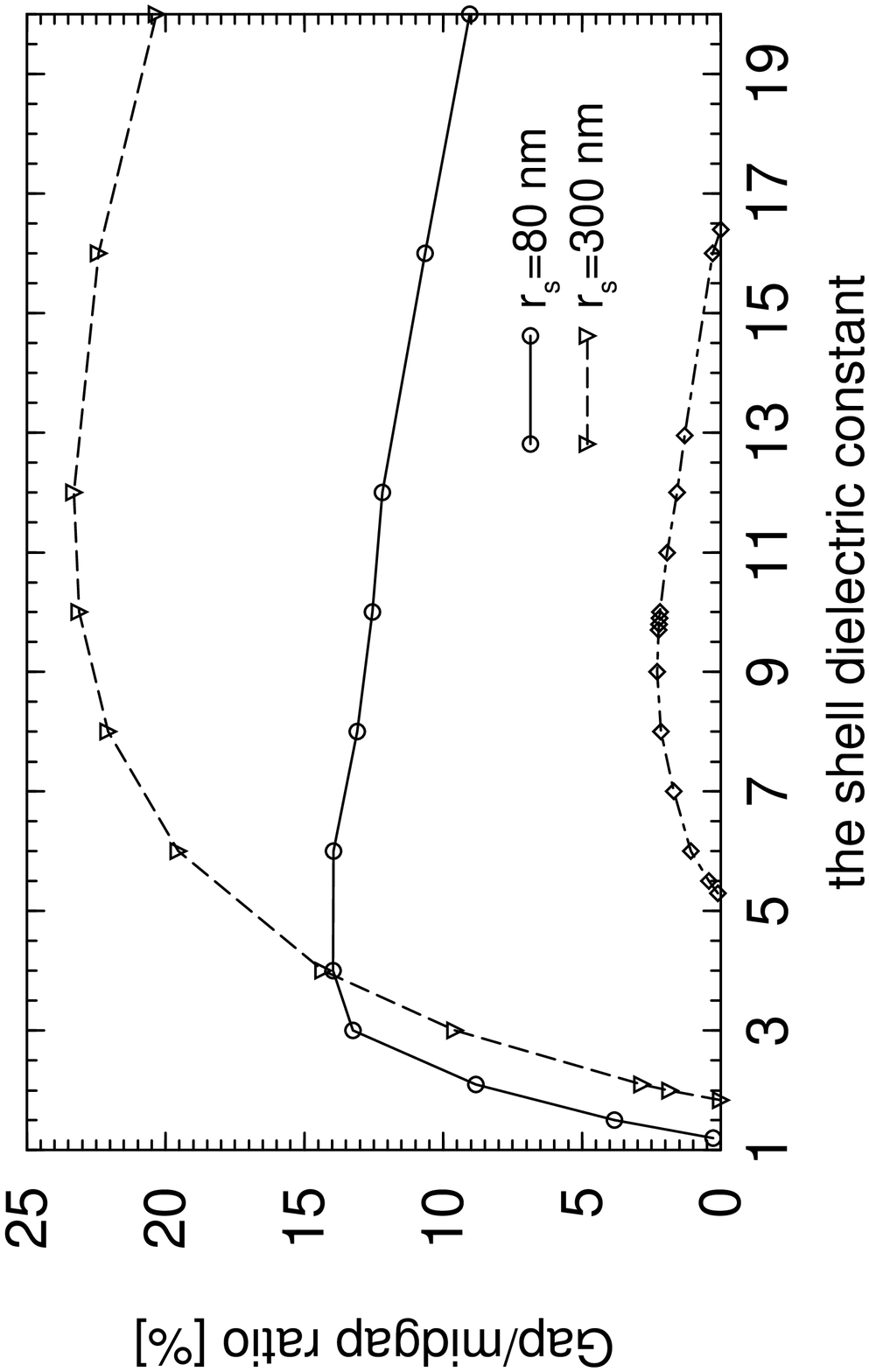,width=12cm,clip=0,angle=0}
\end{center}
\caption{} 
\label{gwdmdrf5eps}
\end{figure}

\begin{figure}[tbp]
\begin{center}
\epsfig{file=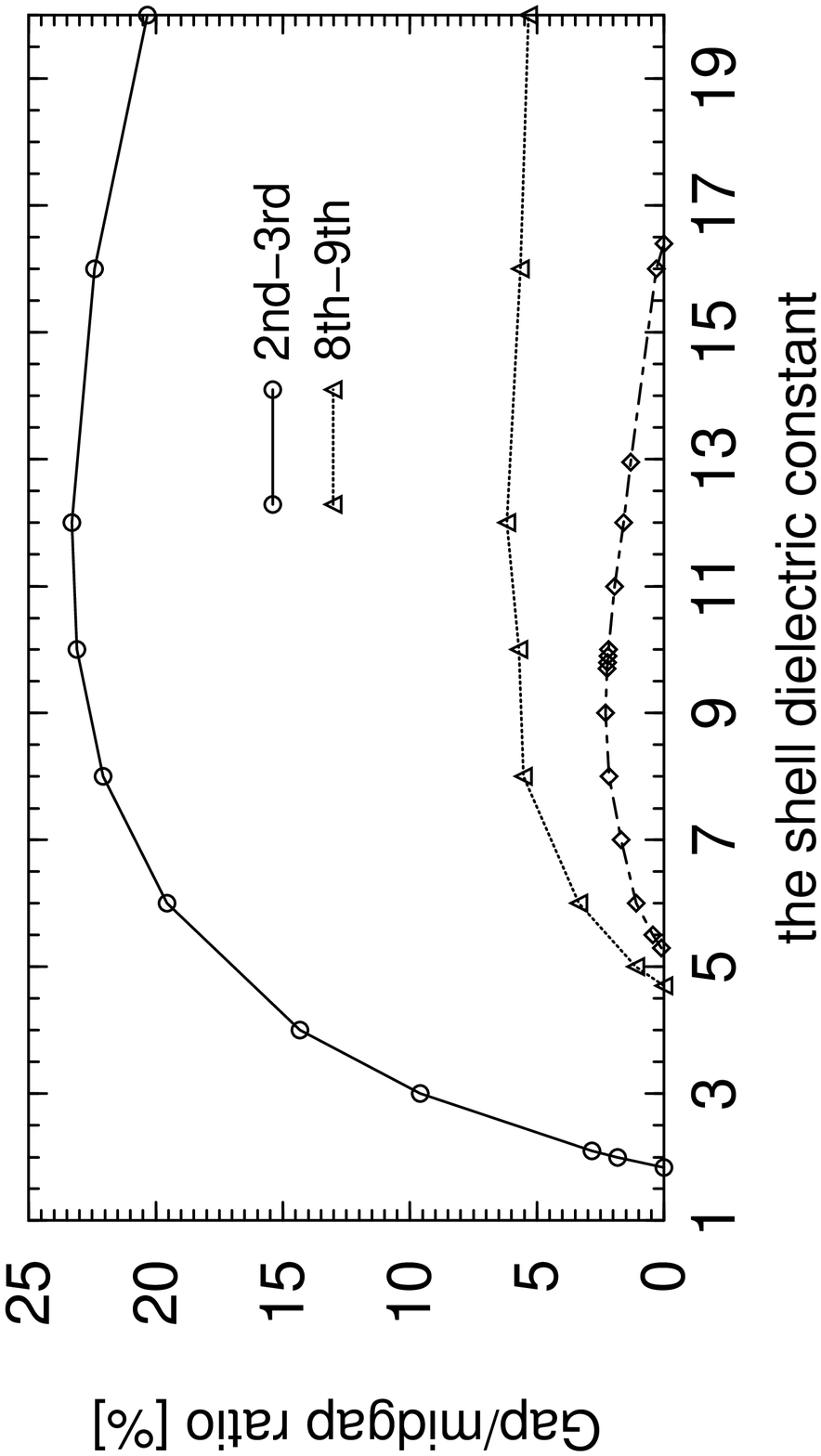,width=12cm,clip=0,angle=0}
\end{center}
\caption{} 
\label{gwdmdrf5r300}
\end{figure}

\begin{figure}[tbp]
\begin{center}
\epsfig{file=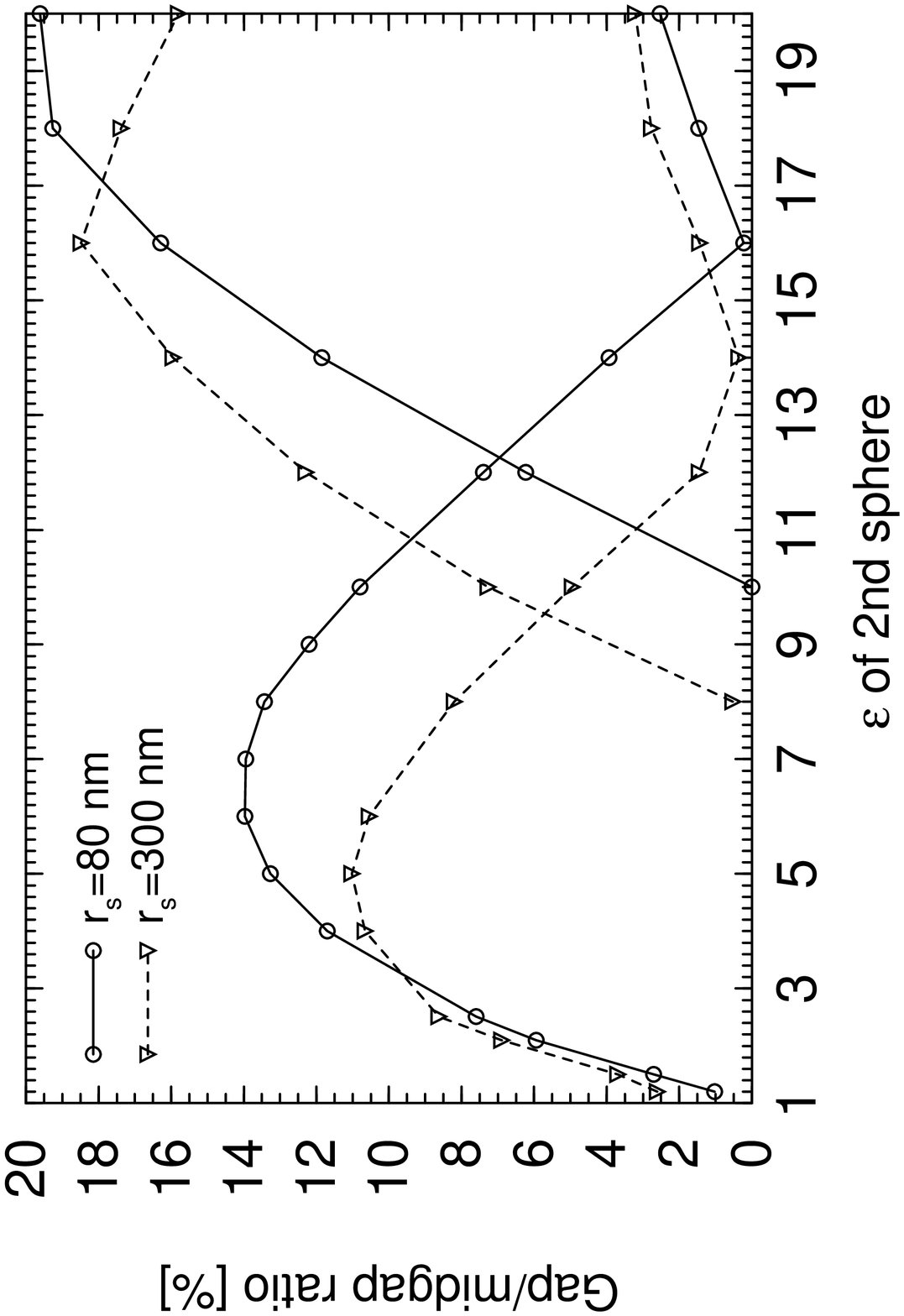,width=12cm,clip=0,angle=0}
\end{center}
\caption{} 
\label{gwT75epsa}
\end{figure}

\begin{figure}[tbp]
\begin{center}
\epsfig{file=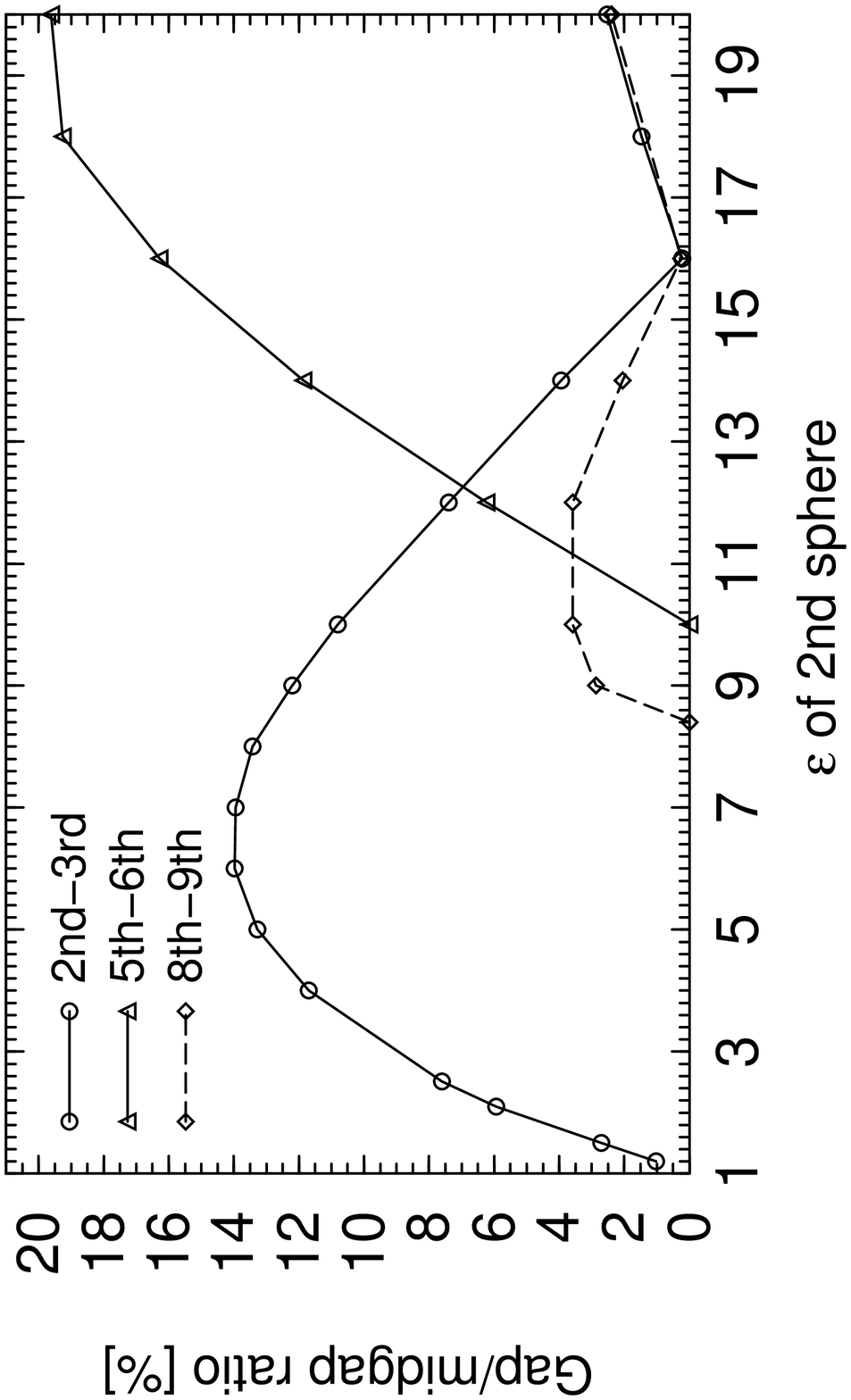,width=12cm,clip=0,angle=0}
\end{center}
\caption{} 
\label{gwr80T75eps}
\end{figure}

\noindent {\bf Figure 3 -}
Convergence of the band structure in Fig. \ref{diamcosta}
as a function of the cutoff $l_{max}$.

\noindent {\bf Figure 4 -}
Photonic band structure for a close-packed zinc-blende lattice of 
spheres in air. One of the two spheres in the primitive lattice cell
has a silver core and dielectric shell with refractive index 
$n_s=1.45$ (silica) with the core to total sphere radii ratio 
$r_c/r_s=0.75$ ($f_m=7.2\%$), whereas the other is a homogeneous 
dielectric sphere with $\varepsilon=12$ of the same radius $r_s=80$ nm. 
Frequency is in units [c/A], as in Fig. \ref{diamcosta}.

\noindent {\bf Figure 5 -}
Calculated gap width  to midgap frequency ratio of 
the $2$nd-$3$rd CPBG for a
close-packed  diamond lattice of dielectric
$n_s=1.45$ (silica) and $n_s=2$ (ZnS)
coated silver spheres in air. The upper graph is for
the sphere radius $r_s=80$ nm and the lower graph is for
the sphere radius $r_s=300$ nm. 
The gap to midgap ratio is plotted as a
function of the metal core radial filling fraction $r_c/r_s$. 
Metal volume fraction is then $f_m=0.34\times (r_c/r_s)^3$.

\noindent {\bf Figure 6 -}
Calculated gap width  to midgap frequency ratio of 
the $2$nd-$3$rd CPBG for a  close-packed metallo-dielectric 
zinc-blende lattice of spheres 
in air. Lattice primitive cell contains a silica sphere and 
a silver core -- $n_s=1.45$ (silica) shell sphere of the same radii. 
$g_w$ is plotted as a function of the ratio $r_c/r_s$ of 
the silver core -- $n_s=1.45$ (silica)
shell sphere
for the cases $r_s=80$ nm and $r_s=300$ nm. 
Metal-volume fraction is then $f_m=0.17\times (r_c/r_s)^3$.
 
\noindent {\bf Figure 7 -}
Calculated gap width to midgap frequency ratio of the 
$2$nd-$3$rd CPBG for a zinc-blende lattice of spheres in air 
for the sphere radius of $r_s=300$ nm. One of the two 
spheres in the lattice primitive cell is a 
silver core -- $n_s=1.45$ (silica) shell sphere, whereas 
the other is a dielectric sphere of the same radius.
Figure shows the effect of the second sphere on the 
$2$nd-$3$rd CPBG which is either a 
$n_c=2$ (ZnS) core -- $n_s=1.45$ (silica) shell
dielectric sphere with fixed $r_c/r_s=0.60$ (dashed line) 
-- or a homogeneous $n=1.45$ (silica) sphere (solid line).

\noindent {\bf Figure 8 -}
Gap width  to midgap frequency ratio of the $2$nd-$3$rd CPBG for a  
close-packed metallo-dielectric diamond lattice of silver core
- dielectric shell spheres with $r_c/r_s=0.5$ ($f_m=4.25\%$) 
in air as a function of the shell dielectric constant
for the cases $r_s=80$ nm and $r_s=300$ nm. For a comparison, 
the dot-dashed line with diamonds shows the gap to midgap 
frequency ratio of the $2$nd-$3$rd CPBG 
of a close-packed diamond lattice of purely dielectric spheres.

\noindent {\bf Figure 9 -}
Gap width  to midgap frequency ratio of the $2$nd-$3$rd 
and $8$th-$9$th CPBG's for a close-packed metallo-dielectric 
diamond lattice of silver core
-- dielectric shell spheres with $r_c/r_s=0.5$ ($f_m=4.25\%$) 
in air as a function of the shell dielectric constant for 
$r_s=300$ nm. For a comparison, the dot-dashed line with 
diamonds shows the gap to midgap frequency ratio of 
the $2$nd-$3$rd CPBG of a close-packed diamond lattice of 
purely dielectric spheres.

\noindent {\bf Figure 10 -}
Calculated gap width  to midgap frequency ratio of 
the $2$nd-$3$rd and the $5$th-$6$th CPBG's for a  
close-packed metallo-dielectric zinc-blende lattice of 
spheres in air. One of the two spheres in the primitive lattice cell
is a silver core -- $n_s=1.45$ (silica) shell sphere with
$r_c/r_s=0.75$-- whereas the other is a homogeneous 
dielectric sphere of the same radius $r_s$ ($f_m=7.2\%$). 
The gap/midgap ratio is plotted as a
function of the dielectric constant of the homogeneous
sphere for the cases $r_s=80$ nm and $r_s=300$ nm. With increasing
dielectric constant of the homogeneous sphere, first the $2$nd-$3$rd 
CPBG opens. The appearance of the $5$th-$6$th CPBG roughly coincides
with the point after which the $2$nd-$3$rd 
CPBG begins to narrow.

\noindent {\bf Figure 11 -}
Calculated gap width  to midgap frequency ratio of 
the $2$nd-$3$rd, $5$th-$6$th, and $8$th-$9$th CPBG's for a  
close-packed metallo-dielectric zinc-blende lattice of spheres in air. 
One of the two spheres in the primitive lattice cell
is a silver core -- $n_s=1.45$ (silica) shell sphere with
$r_c/r_s=0.75$-- whereas the other is a homogeneous 
dielectric sphere of the same radius $r_s=80$ nm ($f_m=7.2\%$). 
The gap/midgap ratio is plotted as a function of the dielectric constant 
of the homogeneous sphere. Note that around $\varepsilon\approx 16$ only 
the $5$th-$6$th CPBG remains open.

\noindent {\bf Figure 12 -}
Calculated gap width  to midgap frequency ratio of the $2$nd-$3$rd CPBG for a  
close-packed metallo-dielectric zinc-blende lattice of spheres in air
as a function of the sphere radius. One of the two spheres in the primitive 
lattice cell is a silver core -- $n_s=1.45$ (silica) shell sphere with
$r_c/r_s=0.75$-- whereas the other is a dielectric sphere of the same 
radius $r_s$ ($f_m=7.2\%$). The dashed line corresponds to the 
case when the 2nd sphere is a homogeneous $n=2$ (ZnS) sphere, the 
solid line is the case of a $n_c=2$ (ZnS) core -- $n_s=1.45$ (silica)
shell sphere with  fixed $r_c/r_s=0.60$, and the dotted line is for 
the case of a homogeneous $n=1.45$ (silica) sphere.

\begin{figure}[tbp]
\begin{center}
\epsfig{file=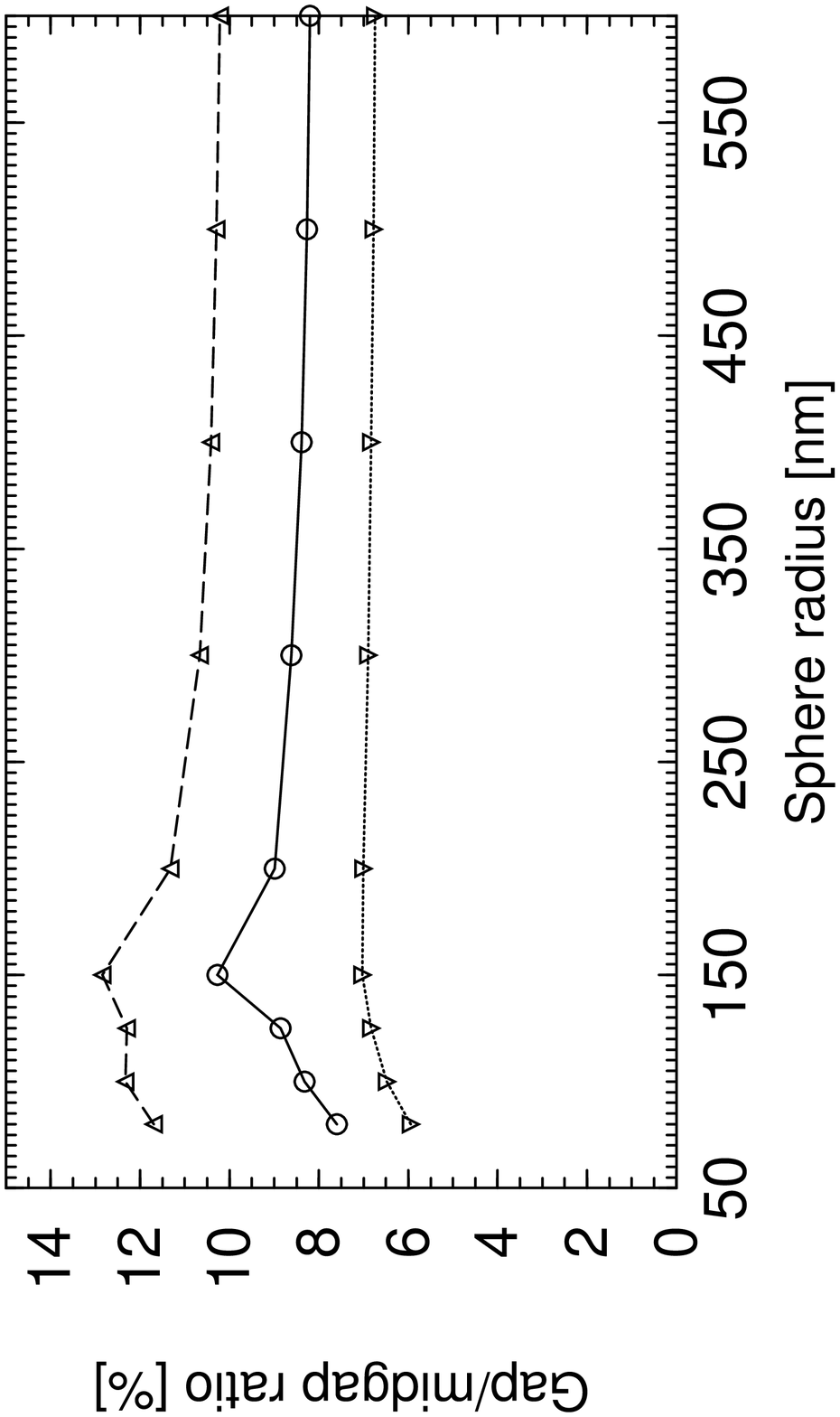,width=12cm,clip=0,angle=0}
\end{center}
\caption{} 
\label{gwT75r}
\end{figure}

\noindent {\bf Figure 13 -}
An example of scaling of the midgap wavelength
(solid line) of the $2$nd-$3$rd CPBG for a close-packed zinc-blende lattice 
of spheres in air with the sphere radius. Gap edges are plotted by dashed lines.
One of the two spheres in the lattice primitive cell is a silver 
core -- $n_s=1.45$ (silica) shell sphere with fixed $r_c/r_s=0.75$ -- 
and the other sphere is a homogeneous $n=1.45$ (silica) sphere of the 
same radius ($f_m=7.2\%$). Gap width  to midgap frequency 
ratio of the $2$nd-$3$rd 
CPBG of the structure is shown by dotted line in Fig. \ref{gwT75r}.

\noindent {\bf Figure 14 -}
An example of scaling of the midgap wavelength of the $2$nd-$3$rd CPBG for
a close-packed zinc-blende lattice of spheres in air with the 
sphere radius. One of the two spheres in the lattice primitive
cell is $n_s=1.45$ (silica) shell sphere with a silver core  
with fixed $r_c/r_s=0.75$, whereas the other is a purely 
dielectric sphere of the same radius ($f_m=7.2\%$). The latter is either 
a homogeneous $n=1.45$ (silica) sphere, a homogeneous 
$n=2$ (ZnS) sphere, or a $n_c=2$ (ZnS) core--$n_s=1.45$ (silica) shell
sphere with fixed $r_c/r_s=0.6$. Gap width  to midgap frequency ratio of 
the $2$nd-$3$rd CPBG of these structures is shown in Fig. \ref{gwT75r}.

\noindent {\bf Figure 15 -}
The extinction efficiency for a silver core -- $n_s=1.45$ (silica) shell 
sphere with fixed $r_c/r_s=0.75$ for different sphere radii.

\noindent {\bf Figure 16 -} 
An example of scaling of the midgap wavelength of the $2$nd-$3$rd CPBG for
a close-packed diamond lattice of coated silver spheres in air with the 
sphere radius. Spheres are either $n_s=1.45$ (silica) coated silver 
spheres with fixed
$r_c/r_s=0.6$, or  $n_s=2$ (ZnS) coated silver 
spheres with fixed $r_c/r_s=0.4$.

\begin{figure}[tbp]
\begin{center}
\epsfig{file=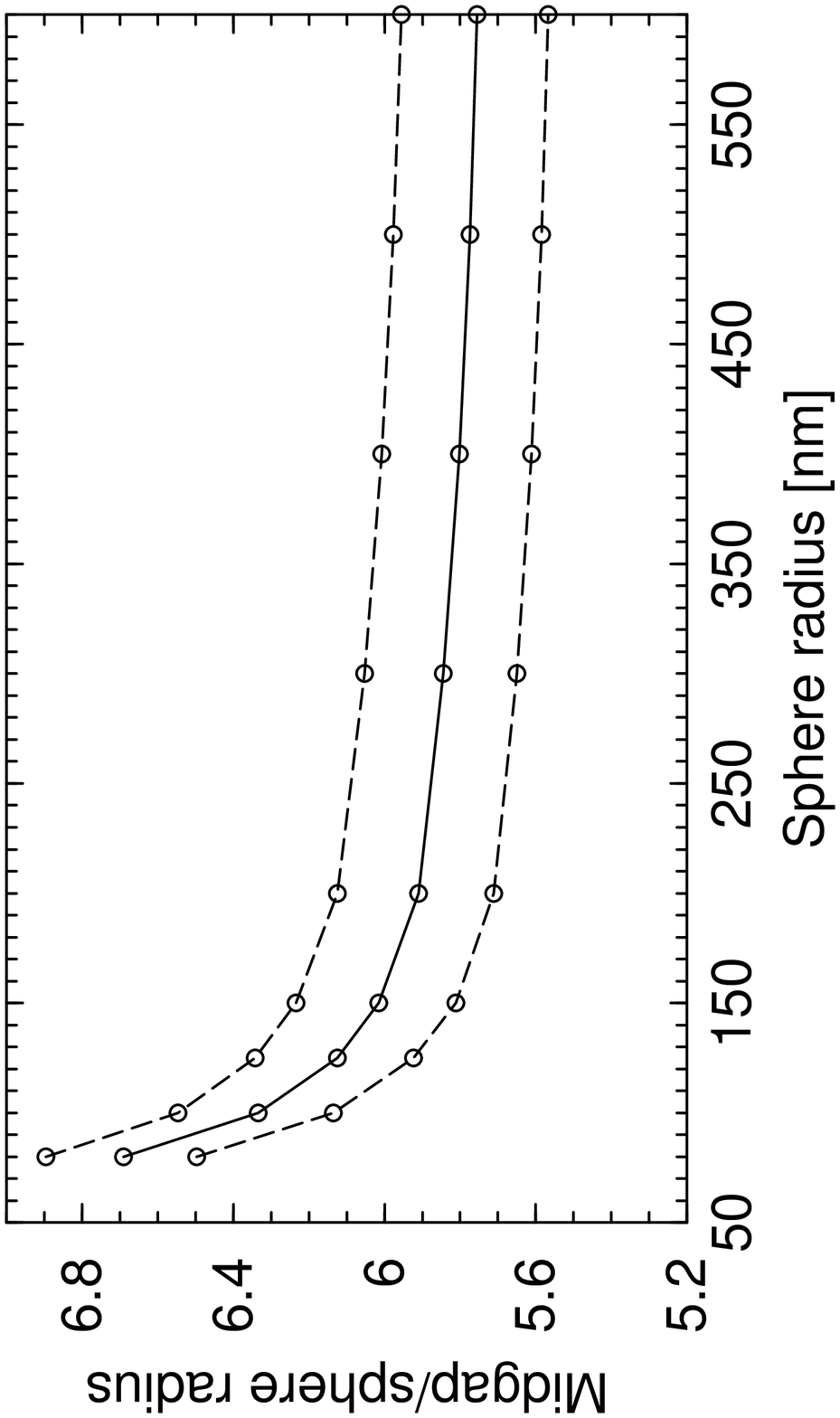,width=12cm,clip=0,angle=0}
\end{center}
\caption{} 
\label{midgapsiocpT75r}
\end{figure}

\begin{figure}[tbp]
\begin{center}
\epsfig{file=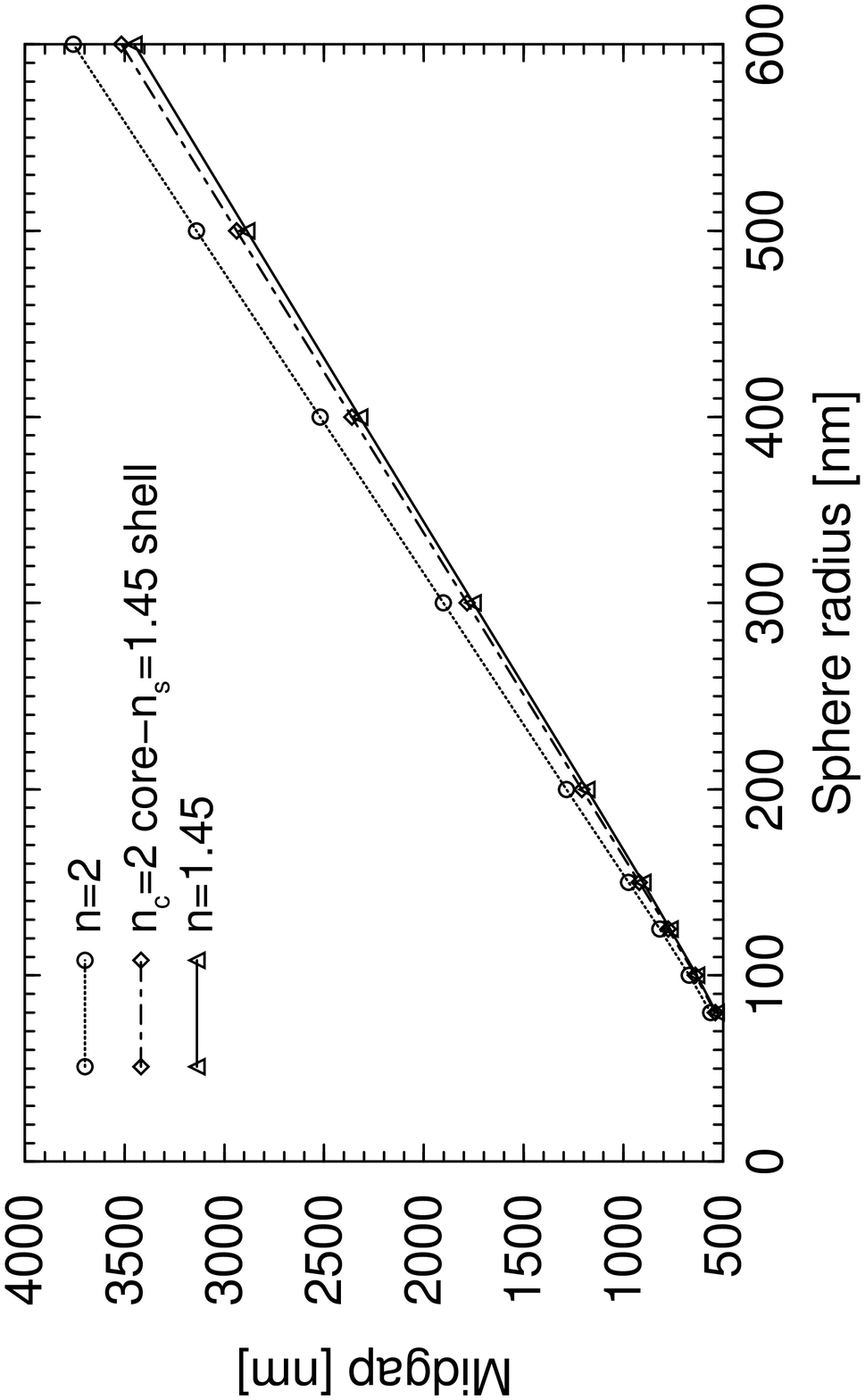,width=12cm,clip=0,angle=0}
\end{center}
\caption{} 
\label{midgapT75}
\end{figure}

\begin{figure}[tbp]
\begin{center}
\epsfig{file=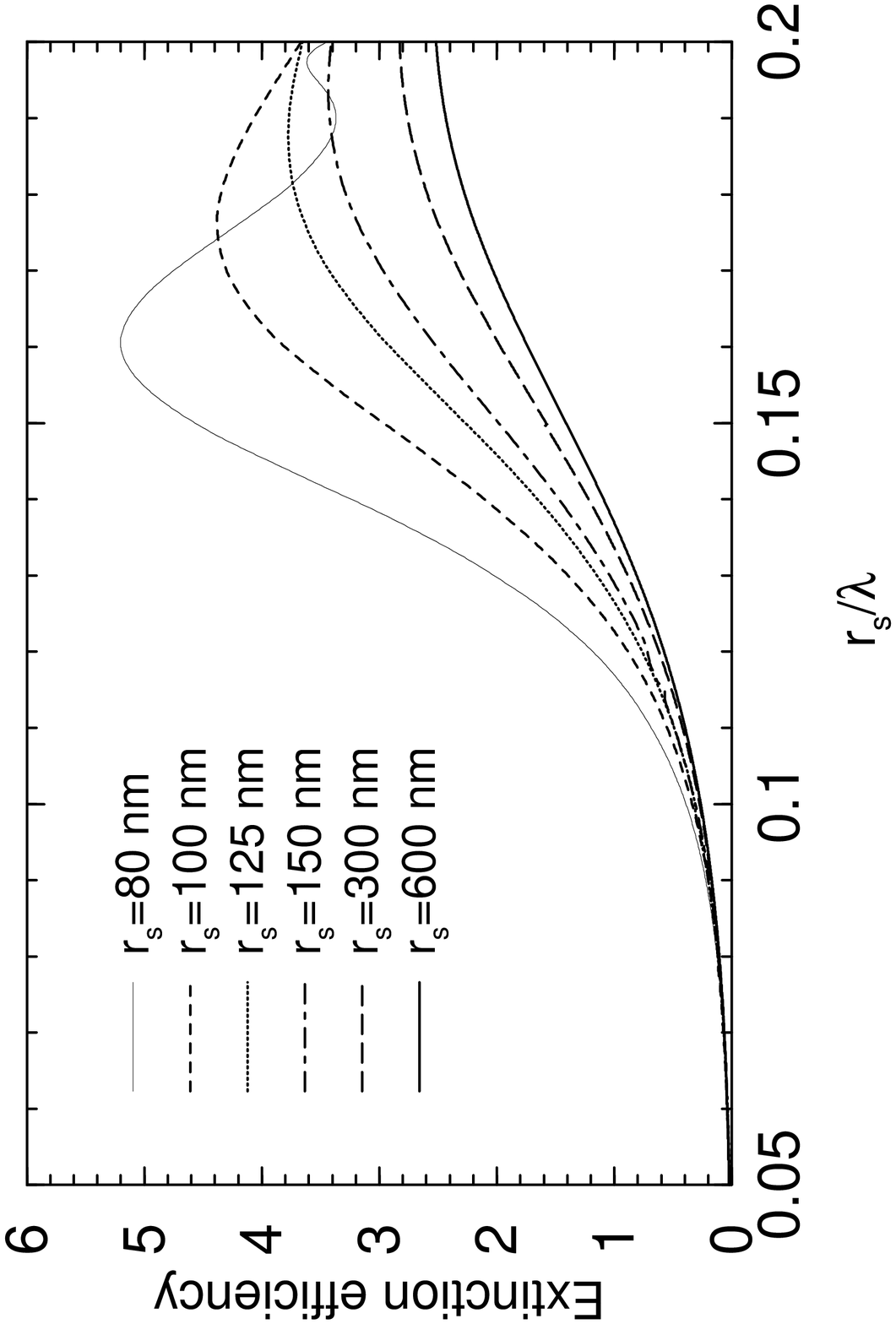,width=12cm,clip=0,angle=0}
\end{center}
\caption{} 
\label{ssextT75sior}
\end{figure}

\begin{figure}[tbp]
\begin{center}
\epsfig{file=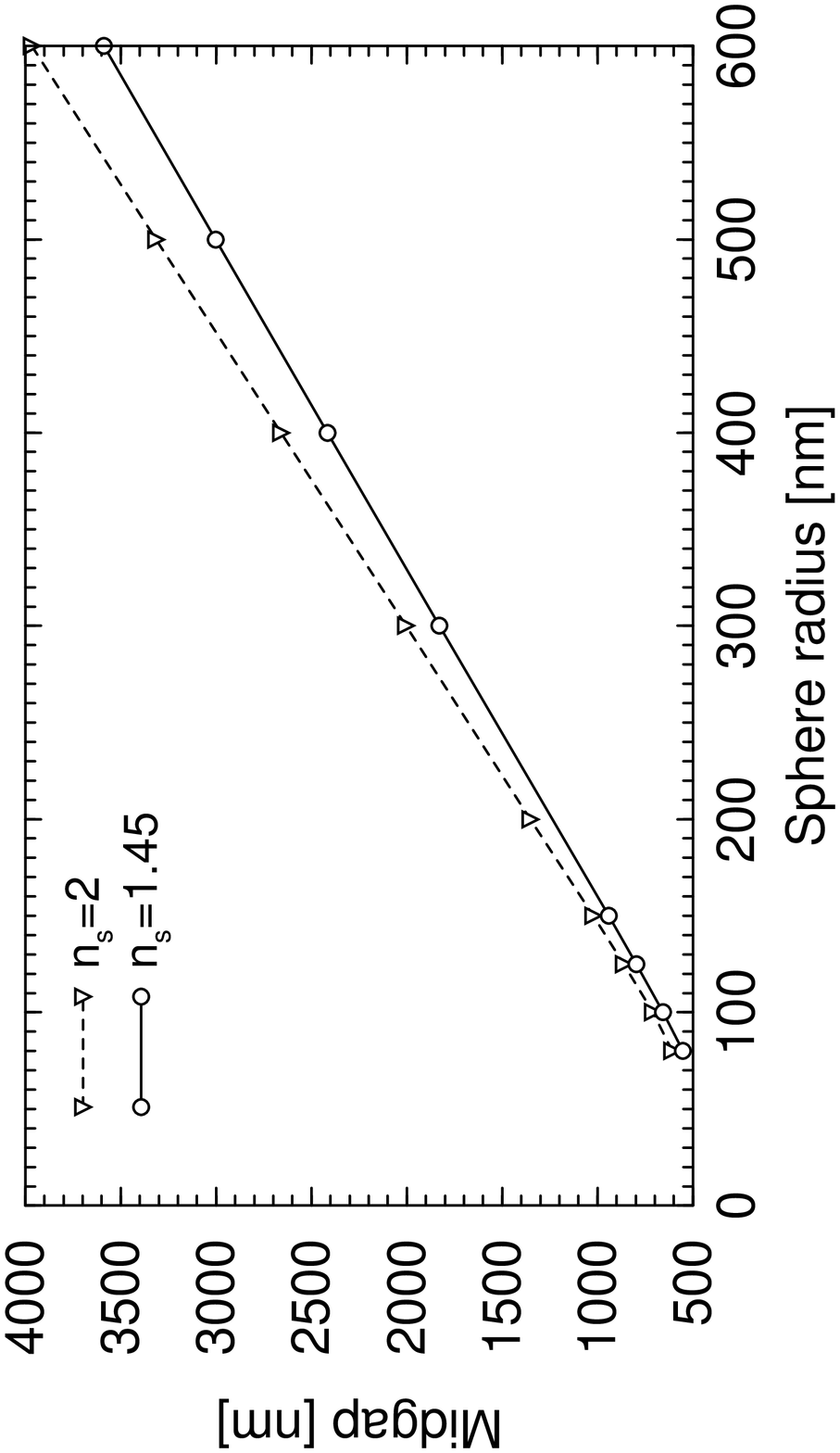,width=12cm,clip=0,angle=0}
\end{center}
\caption{} 
\label{midgapsio}
\end{figure}

\noindent {\bf Figure 17 -}
Reflectance, transmittance, and absorptance
of light incident normally on a two unit cells
(12 planes) thick zinc-blende lattice of spheres in 
air stacked in the (111) direction. The spheres are as in Figs.
\ref{gwT75r} and \ref{midgapsiocpT75r}. Dimensionless frequency
is used on the $x$ axis, where $A$ is the lattice constant
of a conventional unit cell of the cubic lattice.
In all cases, the $2$nd-$3$rd CPBG lies between $\approx$ 0.7 and 0.8.
Gap width  to midgap frequency ratio of 
the $2$nd-$3$rd CPBG of these structures is shown in Fig. \ref{gwT75r},
the midgap wavelength of the $2$nd-$3$rd CPBG is shown in 
Figs. \ref{midgapsiocpT75r} and \ref{midgapT75}.

\noindent {\bf Figure 18 -}
AABBCC stacking of our diamond and zinc-blende structures in 
the (111) direction. Stacking is along the
body diagonal of the conventional unit cell of the cubic lattice.
Lines perpendicular to the stacking $z$ direction indicate hexagonal 
planes of spheres with the primitive lattice vectors 
${\bf a}_1$ and ${\bf a}_2$. Nearest-neighbor distance
of hexagonal planes depicted by the same line
is $1/3$ of the body diagonal.
Spheres in the planes separated by $1/4$ of the body diagonal
have identical $xy$ positions. Spheres in the 
planes separated by $1/12$ of the body diagonal
are shifted by  $({\bf a}_1+{\bf a}_2)/3$ with respect to each other.
The close-packed radius is $1/8$ of the body diagonal.

\begin{figure}[tbp]
\begin{center}
\epsfig{file=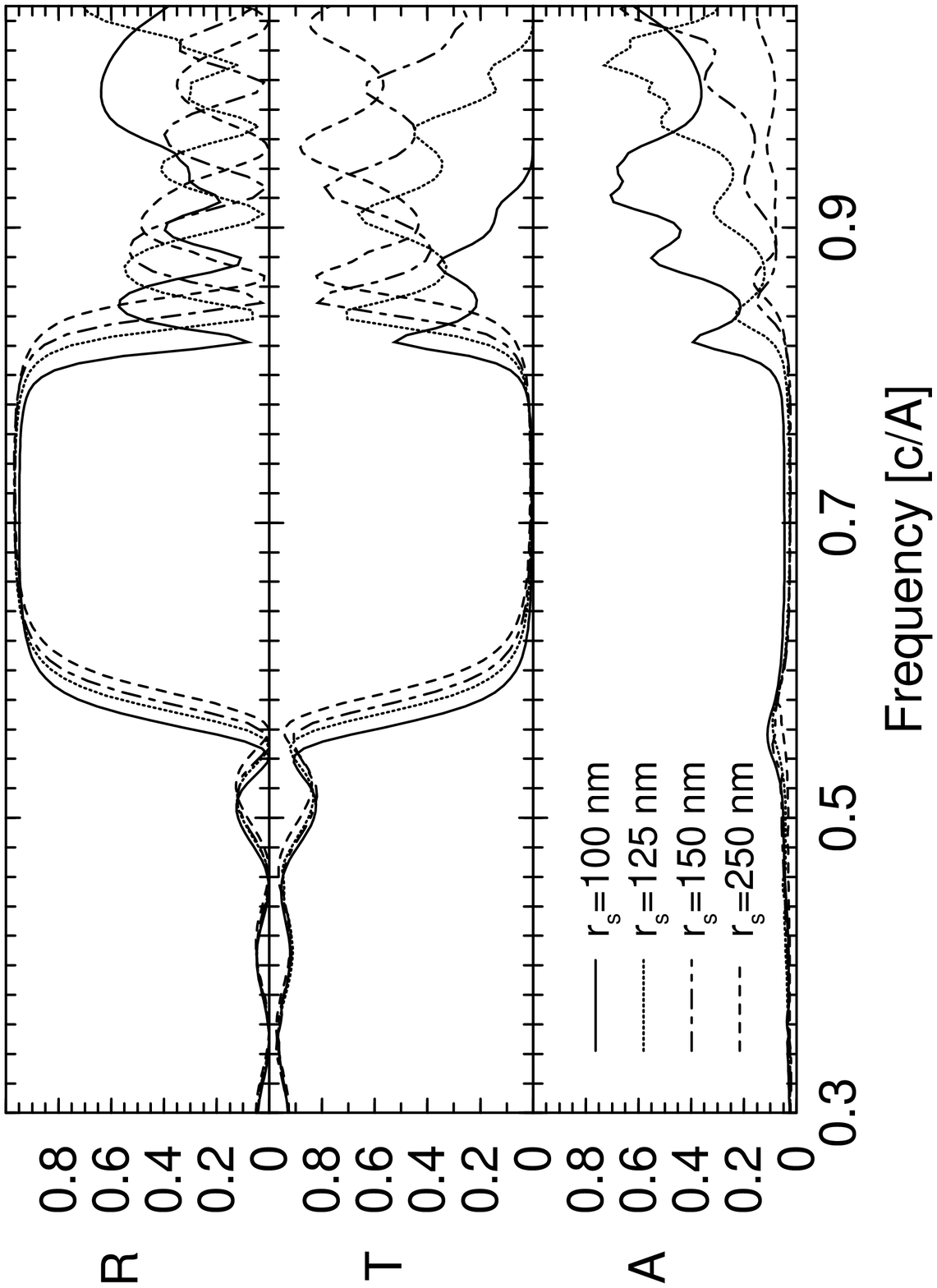,width=12cm,clip=0,angle=0}
\end{center}
\caption{} 
\label{dmdsiocpT75RTA}
\end{figure}

\begin{figure}[tbp]
\begin{center}
\epsfig{file=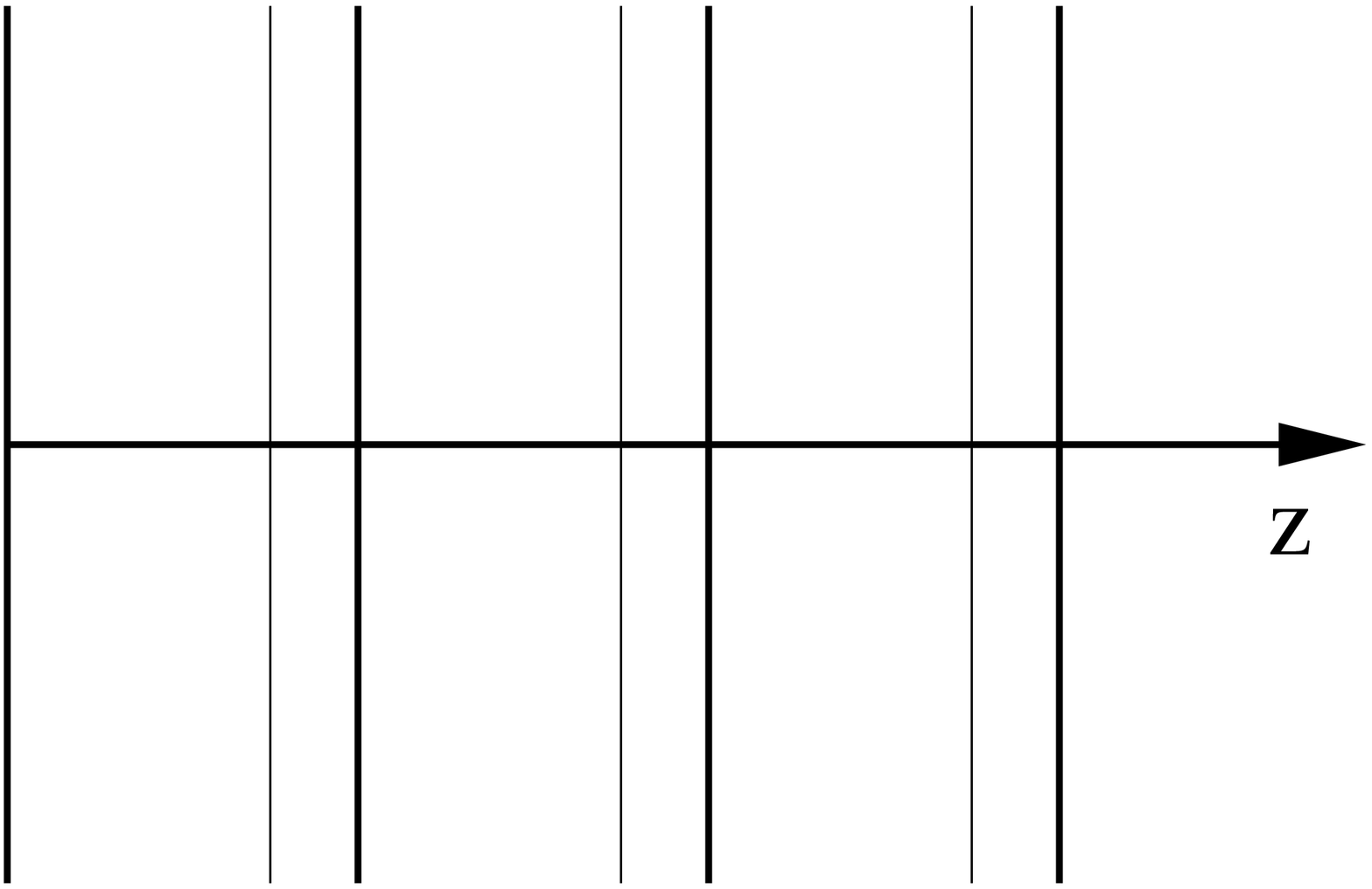,width=12cm,clip=0,angle=0}
\end{center}
\caption{} 
\label{dmd111stack}
\end{figure}

\noindent {\bf Figure 19 -}
Reflectance, transmittance, and absorptance
of light incident normally on a two-unit-cell 
(12 planes) thick zinc-blende lattice of spheres in 
air stacked in the (111) direction as a function of frequency
for geometrically identical cores made of different metals. 
The geometry of $r_s=250$ nm spheres is the same as in 
Fig. \ref{dmdsiocpT75RTA}. Dimensionless frequency
is used on the $x$ axis, where $A$ is the lattice constant
of a conventional unit cell of the cubic lattice.
In all cases, the $2$nd-$3$rd CPBG lies between $\approx$ 0.7 and 0.8.

\begin{figure}[tbp]
\begin{center}
\epsfig{file=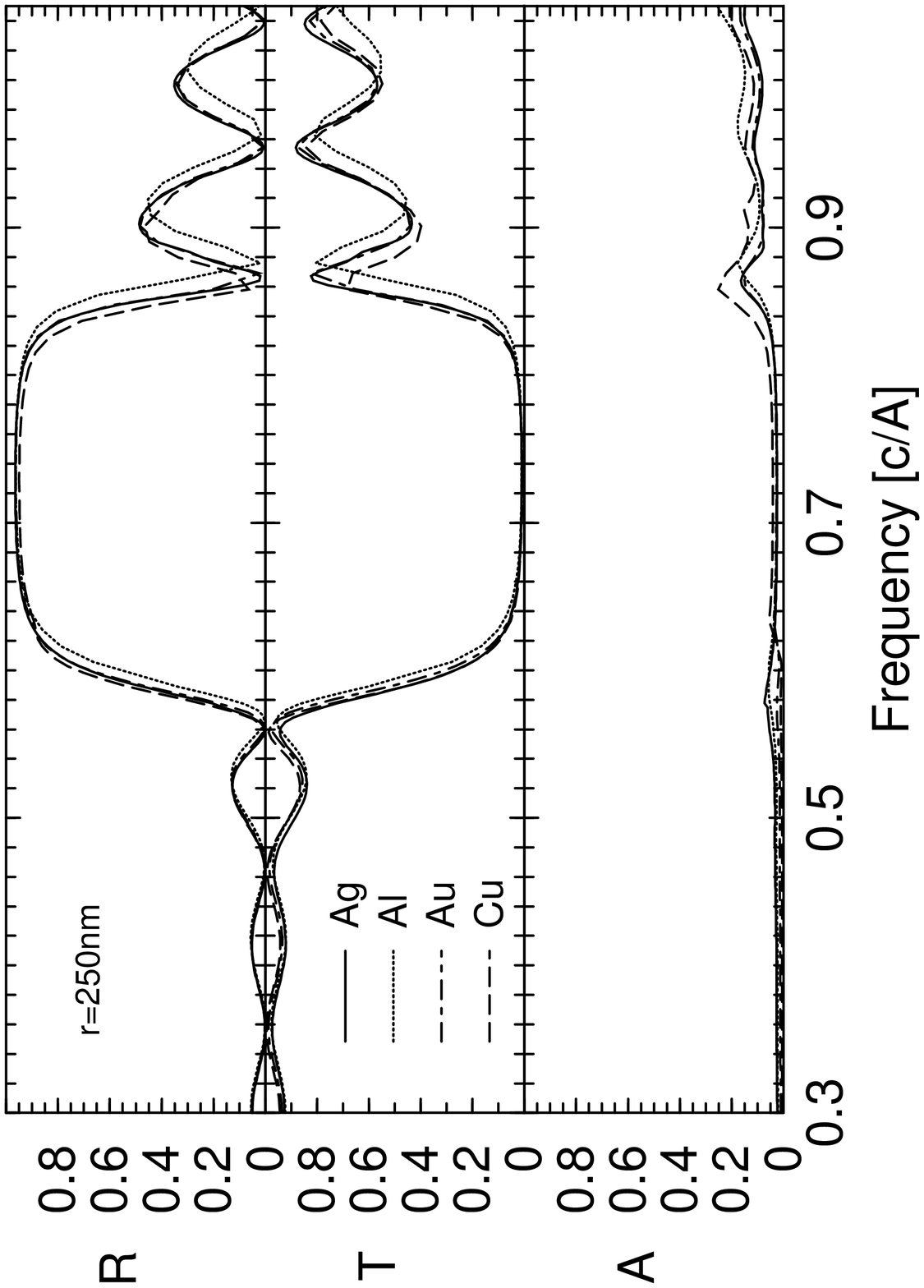,width=12cm,clip=0,angle=0}
\end{center}
\caption{} 
\label{dmdsiocprT75RTA250}
\end{figure}

\begin{figure}[tbp]
\begin{center}
\epsfig{file=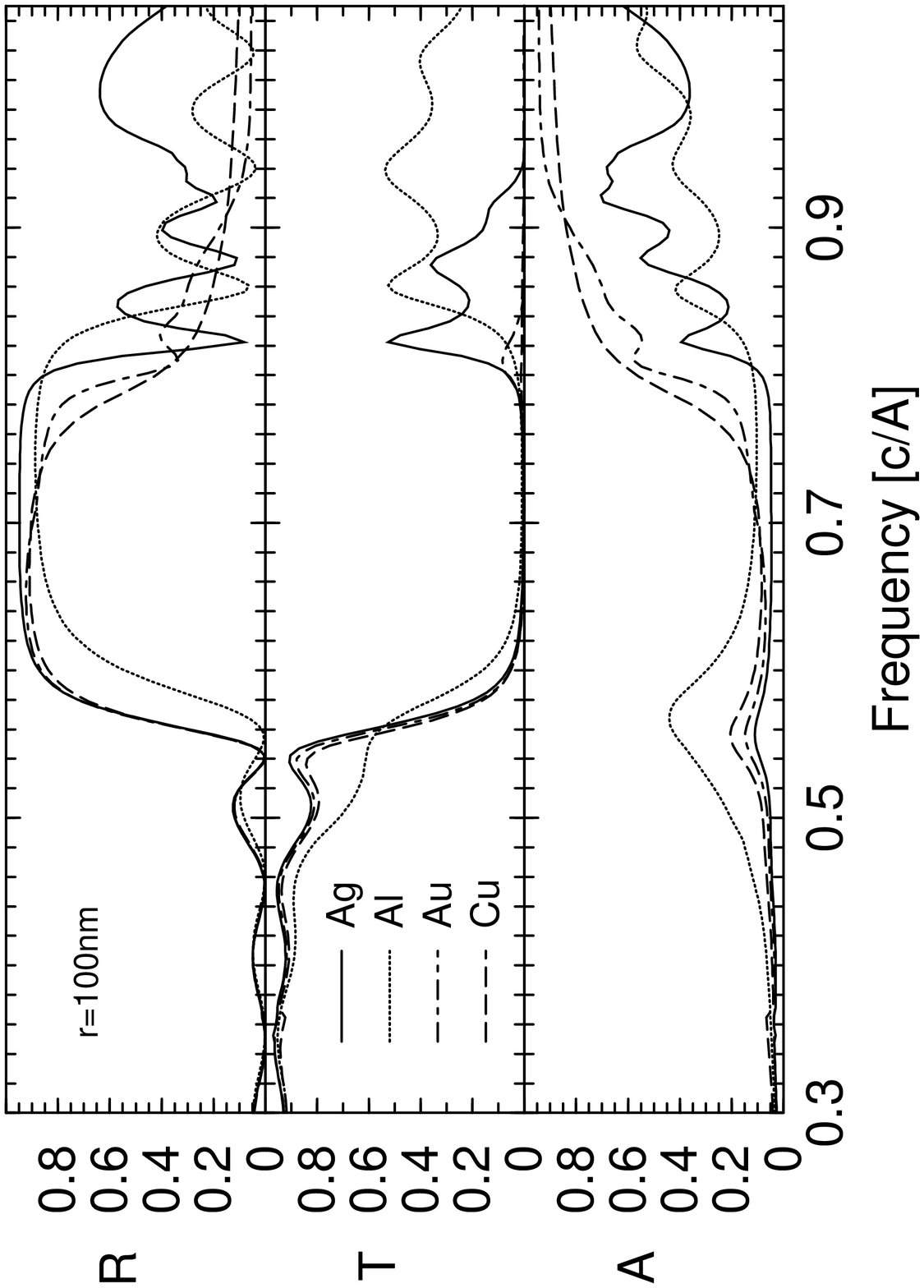,width=12cm,clip=0,angle=0}
\end{center}
\caption{} 
\label{dmdsiocprT75RTA100}
\end{figure}

\begin{figure}[tbp]
\begin{center}
\epsfig{file=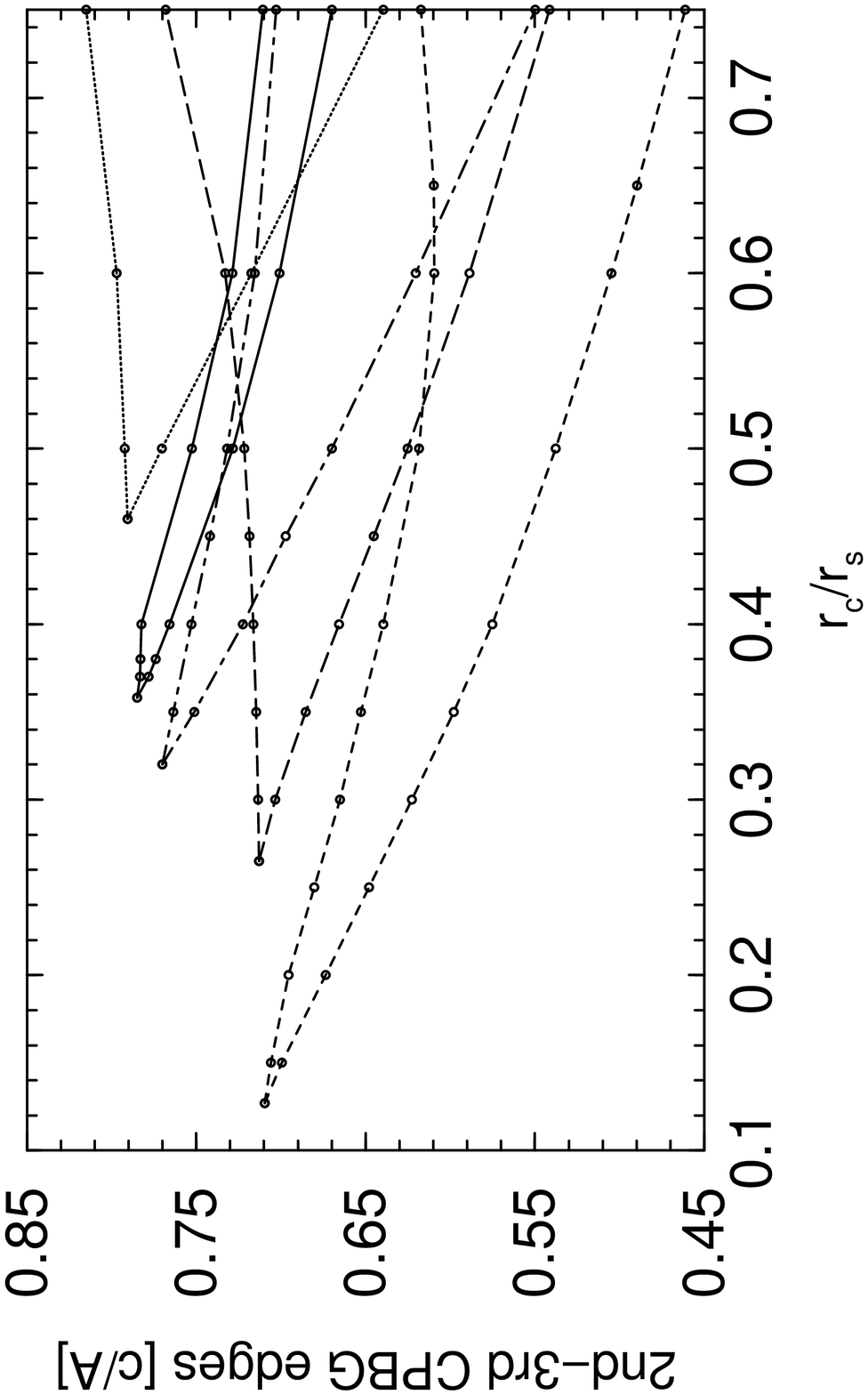,width=12cm,clip=0,angle=0}
\end{center}
\caption{} 
\label{edgesxx}
\end{figure}

\begin{figure}[tbp]
\begin{center}
\epsfig{file=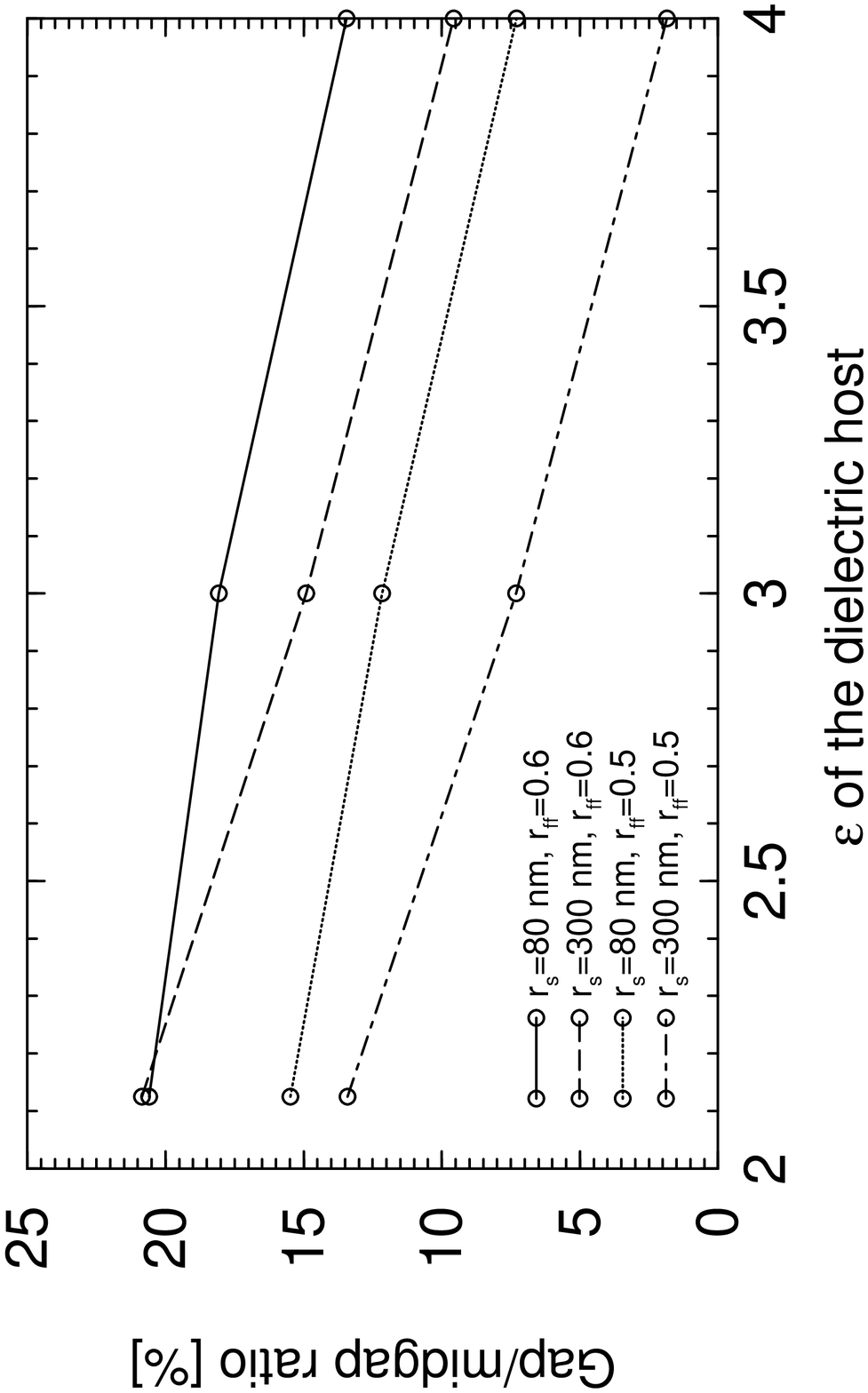,width=12cm,clip=0,angle=0}
\end{center}
\caption{} 
\label{gwepsc8eps}
\end{figure}

\begin{figure}[tbp]
\begin{center}
\epsfig{file=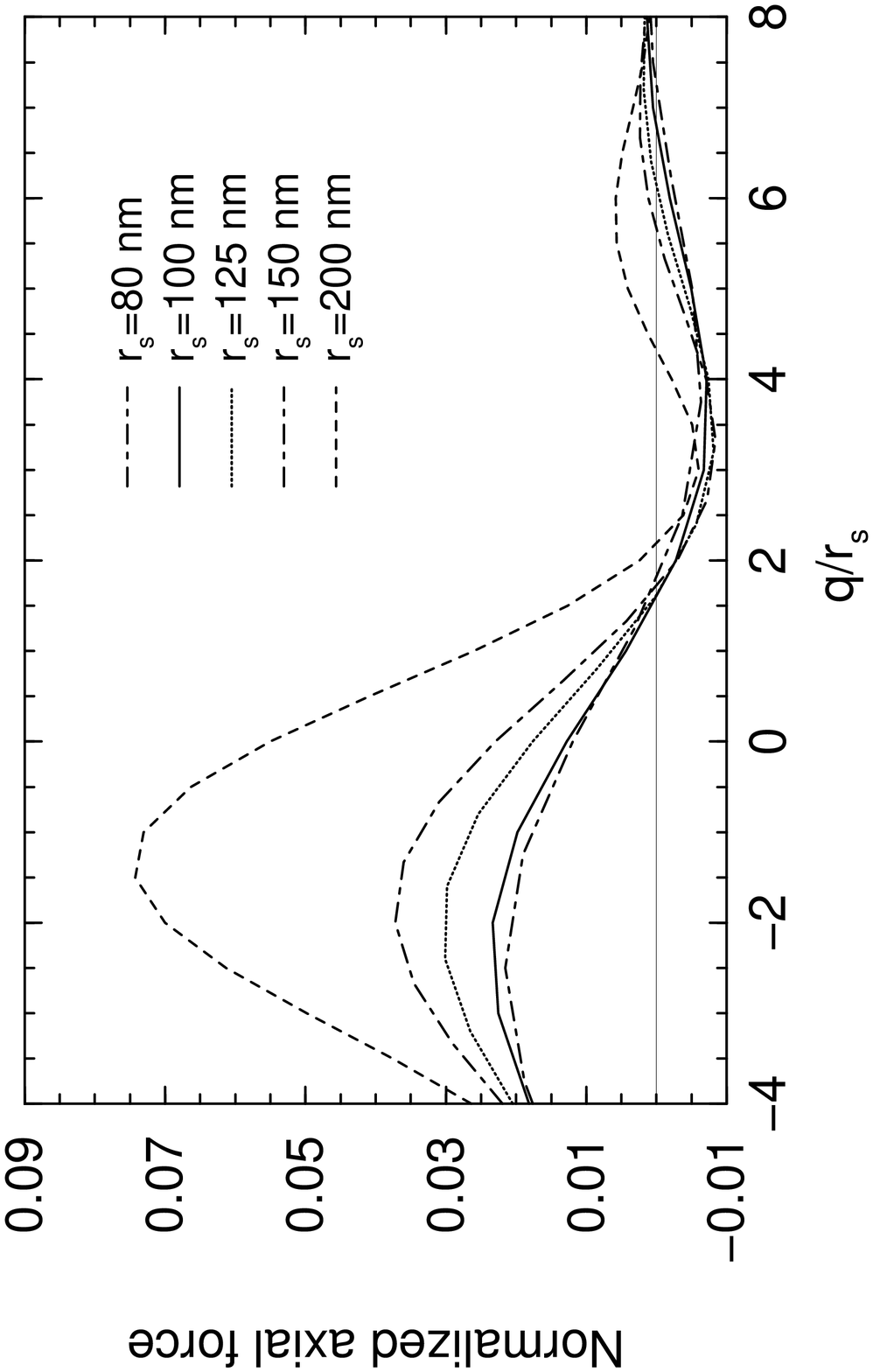,width=12cm,clip=0,angle=0}
\end{center}
\caption{} 
\label{opttraprf3}
\end{figure}

\noindent {\bf Figure 20 -}
The same as in Fig. \ref{dmdsiocprT75RTA250} except that the
sphere radius is now $100$ nm.

\noindent {\bf Figure  21 -}
Behavior of the $2$nd-$3$rd CPBG edges with an increasing relative
radius of the silver core, $r_c/r_s$. Shown are the cases of a
zinc-blende lattice having Ag@SiO${}_2$ and SiO${}_2$ spheres of
the same radius $r_s=80$ nm in the primitive cell (solid lines), a
diamond lattice of silver-core -- $n_s=1.45$-shell spheres
with radii $r_s=80$ nm (dot-dashed lines) and $r_s=300$ nm 
(dotted lines), and a diamond lattice of silver-core-- $n_s=2$-shell spheres
with radii $r_s=80$ nm (dashed lines) and $r_s=300$ nm 
(long-dashed lines).

\noindent {\bf Figure 22 -}
Calculated gap width  to midgap frequency ratio of the $2$nd-$3$rd CPBG 
for a close-packed metallo-dielectric diamond lattice of spheres as a
function of the host dielectric constant. The dependence is shown for 
the sphere radii $r_s=80$ and $300$ nm. Spheres 
have a silver core and dielectric shell with $\varepsilon=8$.
The core to total sphere radii ratio, $r_{ff}$, is either
$r_c/r_s=0.5$ ($f_m=4.25\%$), or  $r_c/r_s=0.6$ ($f_m=7.34\%$). 
For $r_c/r_s=0.4$ ($f_m=2.2\%$) and the dielectric shell with 
$\varepsilon=4$, the $2$nd-$3$rd CPBG is closed in this range of 
the host dielectric constant.

\noindent {\bf Figure 23 -}
Normalized axial force plotted against $q/r_s$,
the center offset from the focus in units of the sphere radius. 
Silver core -- $n_s=1.45$ (silica) shell spheres with fixed 
$r_c/r_s=0.3$ are immersed in water $n_h=1.33$. Laser 
wavelength $\lambda$ increases  by $100$ nm
from $\lambda=600$ nm for $r_s=80$ nm to 
$\lambda=1000$ nm for $r_s=200$ nm. In all cases a stable
equilibrium is achieved at around $q/r_s\approx 2$.

\newpage

\newpage

\newpage

\begin{references}

\bibitem{By}V. P. Bykov, Sov. Phys. JETP {\bf 35}, 269 (1972); 
Sov. J. Quant. Electron. {\bf 4}, 861 (1975).

\bibitem{Y}E. Yablonovitch,
Phys. Rev. Lett. {\bf 58}, 2059 (1987).
 
\bibitem{Souk}Proceedings of the NATO ASI School
``Photonic Crystals and Localization in the 21st Century", 
edited by C. M. Soukoulis (Kluwer, Dordrecht, 2001).

\bibitem{HCS}K. M. Ho, C. T. Chan, and C. M. Soukoulis, 
Phys. Rev. Lett. {\bf 65}, 3152 (1990).

\bibitem{YGL}E. Yablonovitch, T. J. Gmitter, and K. M. Leung,
Phys. Rev. Lett. {\bf 67}, 2295 (1991).

\bibitem{MCK}A. Mekis, J. C. Chen, I. Kurland, S. Fan,
P. R. Villeneuve, and J. D. Joannopoulos, Phys. Rev. 
Lett. {\bf 77}, 3787 (1996).

\bibitem{MS}A. Moroz and C.  Sommers, 
J. Phys.: Condens. Matter {\bf 11}, 997 (1999); 
A. Moroz, J. Opt. A: Pure Appl. Opt. {\bf 1}, 471 (1999).

\bibitem{AMS}A. Moroz, Photonic Crystals at 
 Near-infrared and Optical Wavelengths, in 
``Organic Optoelectronic Materials, Processing and Devices",
Proceedings of the MRS Fall Meeting 2001, Volume 708.

\bibitem{ZLZ}Z.-Y. Li and Z.-Q. Zhang, Phys. Rev. B {\bf 62}, 
1516 (2000).

\bibitem{Lev}B. G. Levi, Phys. Today, January 1999, p. 17.

\bibitem{WCZ}Z. Wang, C. T. Chan, W. Zhang, N. Ming, and P. Sheng, 
Phys. Rev. B {\bf 64}, 113108 (2001).

\bibitem{AMs}A. Moroz, {\em Towards complete photonic band gap 
structures below infrared wavelengths}, in Ref. \cite{Souk}, 
pp. 373-382.

\bibitem{GAvB}C. Graf and A. van Blaaderen, Langmuir {\bf 18}, 524 (2002).

\bibitem{MGM}L. M. Liz-Marz\'{a}n, M. Giersig, and P. Mulvaney,
Langmuir {\bf 12}, 4329 (1996);
V. V. Hardikar and  E. J. Matijevi\'{c}, 
J. Colloid Interface Sci. {\bf 221}, 133 (2000).

\bibitem{SBMc}S. Simeonov, U. Bass, and A. R. McGurn, 
Physica B {\bf 228}, 245 (1996). 

\bibitem{WZY}X. Wang, X.-G. Zhang, Q. Yu, and B. N. Harmon,
Phys. Rev. B {\bf 47}, 4161 (1993).

\bibitem{Mo}A. Moroz,  Phys. Rev. B {\bf 51}, 2068 (1995).

\bibitem{WM}A. R. Williams and J. van W. Morgan, J. Phys. C: Solid
State {\bf 7}, 37 (1974);

\bibitem{BGZ}W. H. Butler, A. Gonis, and X.-G. Zhang, 
Phys. Rev. B {\bf 45}, 11 527 (1992); {\bf 48}, 2118 (1993). 

\bibitem{MG}J. C. M. Garnett, Philos. Trans. R. Soc. London
{\bf 203}, 385 (1904).

\bibitem{SHI}H. S. S\"{o}z\"{u}er, J. W. Haus, and R. Inguva,
Phys. Rev. B {\bf 45}, 13962 (1992).

\bibitem{JJ}S. G. Johnson and J. D. Joannopoulos, 
Opt. Express {\bf 8}, 173 (2001).

\bibitem{Meg}M. Megens (private communication).

\bibitem{VP}P. R. Villeneuve and M. Pich\'{e}, 
Phys. Rev. B {\bf 46}, 4973 (1992). 

\bibitem{AvB} A. van Blaaderen {\em et al.}, {\em Manipulating Colloidal 
Crystallization for Photonic Applications: from Self-organization to 
do-it-yourself Organization}, in Ref. \cite{Souk}, 
pp. 239-251.

\bibitem{SLM}F. Garc\'{\i}a-Santamaria, C. L\'{o}pez, F. Meseguer, 
F. L\'{o}pez-Tejeira, J. S\'{a}nchez-Dehesa, and H. T. Miyazaki,
Appl. Phys. Lett. {\bf 79}, 2309 (2001).

\bibitem{BRW}A. van Blaaderen, R. Ruel, and P. Wiltzius,
Nature (London) {\bf 385}, 321 (1997). 

\bibitem{AM2}A. Moroz, 
Phys. Rev. Lett. {\bf 83}, 5274 (1999);
Europhys. Lett. {\bf 50}, 466 (2000).

\bibitem{ZLW}W. Y. Zhang, X. Y. Lei, Z. L. Wang, 
D. G. Zheng, W. Y. Tam, C. T. Chan, and P. Sheng, 
Phys. Rev. Lett. {\bf 84}, 2853 (2000).

\bibitem{YSM}V. Yannopapas, N. Stefanou, and A. Modinos,   
Comput. Phys. Commun. {\bf 113}, 49 (1998).

\bibitem{Mod}K. Ohtaka, Phys. Rev. B {\bf 19}, 5057 (1979);
J. Phys. C: Solid St. Phys. {\bf 13}, 667 (1980);
A. Modinos, Physica A {\bf 141}, 575 (1987).

\bibitem{Hand}{\em Handbook of Optical Constants of Solids}, 
edited by E. D. Palik (Academic, New York, 1985). 

\bibitem{OLB}M. A. Ordal,  R. J. Bell, R. W.
Alexander, Jr., L. L. Long, and M. R. Querry, 
Appl. Opt. {\bf 24}, 4493 (1983). 

\bibitem{KSB}I. El-Kady, M. M. Sigalas, R. Biswas, K. M. Ho,
and C. M. Soukoulis, Phys. Rev. B {\bf 62}, 15 299 (2000). 

\bibitem{VMB}K. P. Velikov, A. Moroz, and A. van Blaaderen, 
Appl. Phys. Lett. {\bf 80}, 49 (2002).

\bibitem{TMC}A. Tip, A. Moroz, and J.-M. Combes, 
J. Phys. A: Math. Gen. {\bf 33}, 6223 (2000).

\bibitem{KMP1}V. Kuzmiak and A. A. Maradudin,
Phys. Rev. B {\bf 55} 7427 (1997).

\bibitem{MTC}A. Moroz, A. Tip, and J.-M. Combes, 
Synth. Met. {\bf 116}, 481 (2001).

\bibitem{FVJ}S. Fan, P. R. Villeneuve, and J. D. Joannopoulos, 
Phys. Rev. B {\bf 54}, 11245 (1996).

\bibitem{YMS}V. Yannopapas, A. Modinos, and N. Stefanou, 
Phys. Rev. B {\bf 60}, 5359 (1999).

\bibitem{Zach}W. H. Zachariasen, 
{\em Theory of X-ray Diffraction in Crystals} 
(Dover, New York, 1945).

\bibitem{Ru}P. S. J. Russel, Phys. World, August, 37 (1992).

\bibitem{VCD}K. P. Velikov, Ch. G. Christova, R. Dullens, 
and A. van Blaaderen, Science {\bf 296}, 106 (2002).

\bibitem{HVF}J. P. Hoogenboom, D. L. J. Vossen, 
C. Faivre-Moskalenko, M. Dogterom, and A. van Blaaderen,
Appl. Phys. Lett. {\bf 80}, 4828 (2002).

\bibitem{MNN} P. A. Maia Neto and H. M. Nussenzveig,
Europhys. Lett. {\bf 50}, 702 (2000). 

\bibitem{ZZL}X. Zhang, Z.-Q. Zhang, L.-M. Li, C. Jin, D. Zhang, B.
Man, and B. Cheng, Phys. Rev. B {\bf 61}, 1892 (2000).

\bibitem{FZW}A. N. Fang, W. Y. Zhang, Z. L. Wang, A. Hu, and N. B. Ming,
J. Phys.: Condens. Matter {\bf 13}, 8489 (2001).

\bibitem{ZWHM}W. Y. Zhang, Z. L. Wang, A. Hu, and N. B. Ming,
J. Phys.: Condens. Matter {\bf 12}, 9361 (2000).

\bibitem{OAW}S. J. Oldenburg, R. D. Averitt, S. L. Westcott, 
and N. J. Halas, Chem. Phys. Lett. {\bf 288}, 243 (1998).

\bibitem{MT}A. Moroz and A. Tip, 
J. Phys.: Condens. Matter. {\bf 11}, 2503 (1999).

\bibitem{KR}W. Kohn and N. Rostoker,    
Phys. Rev. {\bf 94},  1111 (1954). 

\bibitem{AS}M. Abramowitch and I. A. Stegun,    
{\em Handbook of Mathematical Functions} 
(Dover Publications, New York, 1973). 

\bibitem{F}J. S. Faulkner,
J. Phys. C: Solid State {\bf 10}, 4661 (1977).

\bibitem{Seg}B. Segall, Phys. Rev. {\bf 105}, 108 (1957).  
 
\end{references}
\end{document}